\documentclass[lettersize,journal]{IEEEtran}

\usepackage{pgf}
\usepackage{tikz}
\usetikzlibrary{arrows}
\usetikzlibrary{automata}
\usetikzlibrary{spy}
\usetikzlibrary{patterns}
\usepackage[utf8]{inputenc}
\usepackage{textcomp}
\usepackage{stfloats}
\usepackage{verbatim}

\usepackage{tabularx,booktabs}
\usepackage{pgfplots}
\usepackage{lipsum}
\usepackage{dsfont}
\usepackage{tikz-layers}

\usepackage{setspace}
\usepackage[ruled,vlined]{algorithm2e}
\pgfplotsset{compat=1.6}
\usepgfplotslibrary{statistics}
\usepgfplotslibrary{colorbrewer} 
\pgfplotsset{compat = 1.15, cycle list/Set1-8} 
\usetikzlibrary{pgfplots.statistics, pgfplots.colorbrewer} 
\usepackage{pgfplotstable}

\usepackage{amsmath,amsthm,amssymb}
\usepackage{optidef}
\usepackage{bm}
\usepackage{mathtools}
\usepackage[cmintegrals]{newtxmath}

\usepackage{glossaries-extra}
\usepackage{stackengine}
\usepackage[euler]{textgreek}
\usepackage{lgreek}
\usepackage[hyphens]{url}
\usepackage{comment}

\usepackage{enumitem}
\usepackage{cite}
\usepackage[hidelinks]{hyperref}
\usepackage{cleveref}
\usepackage{subcaption}
\usepgfplotslibrary{fillbetween}
\usepackage{ifthen}

\newcommand{\mixs}[1]{\texttt{mixServ\!(${#1}$)}}

\newcommand{\su}{\texttt{singleURLLC}}
\newcommand{\mixr}[1]{\texttt{mix-RandSel\!\!(${#1}$)}\!}
\newcommand{\mixo}{\texttt{mix-baseline}}
\newcommand{\slr}[1]{\texttt{slic-RandSel\!\!(${#1}$)}\!}
\newcommand{\slo}{\texttt{slic-baseline}}

\newcommand{\ours}{\texttt{BSAC-Coex}}
\newcommand{\diff}{\,\mathrm{d}}

\newcommand{\bc}{\mbox {\boldmath $c$}}

\newcommand{\bolde}{\mbox {\boldmath $e$}}

\newcommand{\bw}{\mbox {\boldmath $w$}}

\newcommand{\calA}{\mathcal{A}}
\newcommand{\calB}{\mathcal{B}}

\newcommand{\calN}{\mathcal{N}}
\newcommand{\calO}{\mathcal{O}}

\newcommand{\calR}{\mathcal{R}}
\newcommand{\calS}{\mathcal{S}}

\newcommand{\calU}{\mathcal{U}}

\newcommand{\calX}{\mathcal{X}}

\newcommand{\norm}[1]{\left\lVert#1\right\rVert}

\DeclareMathOperator*{\argmax}{arg\,\!max}
\SetKwComment{Comment}{$\triangleright$\ }{}

\newcommand{\R}{\mathbb{R}}
\newcommand{\N}{\mathbb{N}}

\makeatletter
\newcommand\footnoteref[1]{\protected@xdef\@thefnmark{\ref{#1}}\@footnotemark}
\makeatother

\setabbreviationstyle[acronym]{long-short}
\GlsXtrEnableEntryCounting
 {acronym}
 {2}
 
\theoremstyle{definition}
\newtheorem{assumption}{Assumption}

\theoremstyle{definition}

\theoremstyle{remark}

\newcounter{reviewer}
\newcounter{point}[reviewer]
\renewcommand{\thepoint}{\thereviewer.\arabic{point}} 

\newcommand{\hlineB}{\noalign{\hrule height 1pt}}
\definecolor{singleUrllcColor}{rgb}{0.92900,0.69400,0.12500}%
\definecolor{slicingColor}{rgb}{0.71,0.49,0.86}

\definecolor{mixedRandColor1}{rgb}{0.24, 0.2, 0.12}
\definecolor{mixedRandColor2}{rgb}{0.4, 0.69, 0.2}
\definecolor{mixedRandColor3}{rgb}{0,0.27,0.13}
\definecolor{ourSolColor}{rgb}{0.29, 0.59, 0.82}%

\definecolor{reqColor}{rgb}{0.98, 0.38, 0.5}
\definecolor{bgColor}{rgb}{0.82, 0.71, 0.55}

\definecolor{mycolor2}{rgb}{1.0, 0.4, 0.6}
\definecolor{mycolor6}{rgb}{0.43, 0.21, 0.1}

\definecolor{vColor}{rgb}{0.12, 0.3, 0.17}
\definecolor{mycolorOnlyAiOpt}{rgb}{0.12, 0.3, 0.17}
\definecolor{mycolorSlic}{rgb}{0.47, 0.32, 0.66}

\newacronym{TV}{TV}{television}
\newacronym{DTTB}{DTTB}{digital television terrestrial broadcasting}
\newacronym{DVB}{DVB}{Digital Video Broadcast}
\newacronym{DVB-H}{DVB-H}{Digital Video Broadcast-Handheld}
\newacronym{ATSC}{ATSC}{Advanced Television System Committee}
\newacronym{ATSC-M/H}{ATSC-M/H}{Advanced Television System Committee - Mobile/Handheld}
\newacronym{IPTV}{IPTV}{Internet Protocol television}
\newacronym{IP}{IP}{Internet Protocol}

\newacronym{UE}{UE}{user equipment}
\newacronym{PC}{PC}{personal computer}
\newacronym{RAN}{RAN}{radio access network}
\newacronym{CN}{CN}{core network}
\newacronym{MS}{MS}{mobile station}

\newacronym{ITU-R}{ITU-R}{International Telecommunications Union - Radiocommunication Sector}
\newacronym{IMT-Advanced}{IMT-Advanced}{International Mobile Telecommunications Advanced}
\newacronym{4G}{4G}{fourth-generation of mobile phone communications and Internet access technology}

\newacronym{3gpp}{3GPP}{3rd Generation Partnership Project}
\newacronym{GSM}{GSM}{Global System for Mobile Communications}
\newacronym{UMTS}{UMTS}{Universal Mobile Telecommunications System}
\newacronym{HSPA}{HSPA}{High Speed Packet Access}
\newacronym{lte}{LTE}{Long-Term Evolution}
\newacronym{lte-a}{LTE-A}{Long-Term Evolution Advanced}

\newacronym{e-UTRAN}{e-UTRAN}{evolved Universal Terrestrial Radio Access Network}
\newacronym{eNB}{eNB}{e-UTRAN NodeB}
\newacronym{gNB}{gNB}{gNodeB}
\newacronym{EPC}{EPC}{Evolved Packet Core}

\newacronym{MBMS}{MBMS}{Multimedia and Broadcast Multicast Service}
\newacronym{eMBMS}{eMBMS}{Evolved MBMS}
\newacronym{SFN}{SFN}{single-frequency network}
\newacronym{MBSFN}{MBSFN}{MBMS single-frequency network}
\newacronym{BM-SC}{BM-SC}{Broadcast/Multicast Service Center}
\newacronym{MBMS GW}{MBMS GW}{MBMS Gateway}
\newacronym{MME}{MME}{Mobility Management Entity}
\newacronym{MCE}{MCE}{Multi-cell/multicast Coordinating Entity}
\newacronym{SYNC}{SYNC}{synchronization}
\newacronym{MCCH}{MCCH}{Multicast Control Channel}
\newacronym{MTCH}{MTCH}{Multicast Traffic Channel}
\newacronym{MCH}{MCH}{Multicast Channel}
\newacronym{PMCH}{PMCH}{Physical Multicast Channel}
\newacronym{PDSCH}{PDSCH}{Physical Downlink Shared Channel}

\newacronym{IEEE}{IEEE}{Institute of Electrical and Electronics Engineers}
\newacronym{WiMAX}{WiMAX}{Worldwide Interoperability for Microwave Access}
\newacronym{ASN}{ASN}{access service network}
\newacronym{ASN-GW}{ASN-GW}{ASN gateway}
\newacronym{CSN}{CSN}{Connectivity Service Network}

\newacronym{PA}{PA}{power amplifier}

\newacronym{NI}{NI}{National Instruments}

\newacronym{TDD}{TDD}{time-division duplex}
\newacronym{FDD}{FDD}{frequency-division duplex}
\newacronym{UDP}{UDP}{User Datagram Protocol}
\newacronym{APP}{APP}{application}
\newacronym{mac}{MAC}{medium access control}
\newacronym{phy}{PHY}{physical}
\newacronym{RLC}{RLC}{radio link control}
\newacronym{FIFO}{FIFO}{first-in first-out}
\newacronym{CRC}{CRC}{cyclic redundancy check}
\newacronym{SAP}{SAP}{service access point}
\newacronym{FEC}{FEC}{forward error correction}
\newacronym{IF}{IF}{intermediate frequency}
\newacronym{RF}{RF}{radio frequency}
\newacronym{mimo}{MIMO}{multiple-input and multiple-output}
\newacronym{aoa}{AoA}{angle-of-arrival}
\newacronym{aod}{AoD}{angle-of-departure}
\newacronym{mcs}{MCS}{modulation and coding scheme}

\newacronym{SPC}{SPC}{superposition coding}
\newacronym{SVC}{SVC}{Scalable Video Coding}
\newacronym{GM}{GM}{generic multicasting}
\newacronym{SCM}{SCM}{superposition coded multicasting}
\newacronym{SIC}{SIC}{successive interference cancellation}

\newacronym{st}{ST}{secondary transmitter}
\newacronym{pt}{PT}{primary transmitter}
\newacronym{sr}{SR}{secondary receiver}
\newacronym{pr}{PR}{primary receiver}
\newacronym{su}{SU}{secondary user}
\newacronym{pu}{PU}{primary user}

\newacronym{awgn}{AWGN}{additive white Gaussian noise}

\newacronym{pdf}{PDF}{probability density function}
\newacronym{ccdf}{CCDF}{complementary CDF}
\newacronym{iid}{IID}{independent and identically distributed}
\newacronym{rf}{RF}{radio frequency}

\newacronym{dd}{DD}{Device-to-Device}
\newacronym{ddu}{DDU}{Device-to-Device user}
\newacronym{dds}{DDS}{Device-to-Device system}
\newacronym{ddt}{DT}{DDU transmitter}
\newacronym{ddr}{DR}{DDU receiver}

\newacronym{bs}{BS}{base station}
\newacronym{bsu}{BSU}{base station associated user}
\newacronym{bss}{BSS}{base station associated system}
\newacronym{bst}{BT}{BSU transmitter}
\newacronym{bsr}{BR}{BSU receiver}

\newacronym{epg}{EPG}{energy per goodbit}
\newacronym{mepg}{MEPG}{modified energy per goodbit}
\newacronym{ee}{EE}{energy efficiency}
\newacronym{se}{SE}{spectral efficiency}

\newacronym{wrt}{w.r.t.}{with respect to}

\newacronym{kkt}{KKT}{Karush-Kuhn-Tucker}
\newacronym{admm}{ADM}{Alternating Directing Method}
\newacronym{cr}{CR}{cognitive radio}
\newacronym{ssi}{SSI}{soft-sensing information}
\newacronym{csi}{CSI}{channel state information}
\newacronym{qsi}{QSI}{queue state information}
\newacronym{el}{EL}{enhancement layer(s)}
\newacronym{snr}{SNR}{signal-to-noise ratio}

\newacronym{NAL}{NAL}{network abstraction layer}
\newacronym{QP}{QP}{quantization parameter}

\newacronym{ofdma}{OFDMA}{orthogonal frequency-division multiple access}
\newacronym{tdma}{TDMA}{time division multiple access}

\newacronym{PUSC}{PUSC}{partial usage of the subchannels}
\newacronym{CFO}{CFO}{carrier frequency offset}
\newacronym{I/Q}{I/Q}{in-phase and quadrature-phase}
\newacronym{ASK}{ASK}{amplitude-shift keying}
\newacronym{PSK}{PSK}{phase-shift keying}
\newacronym{BPSK}{BPSK}{binary phase-shift keying}
\newacronym{QPSK}{QPSK}{quadrature phase-shift keying}
\newacronym{QAM}{QAM}{quadrature amplitude modulation}
\newacronym{PSNR}{PSNR}{peak signal-to-noise ratio}
\newacronym{PELR}{PELR}{packet error and loss rate}
\newacronym{per}{PER}{packet error ratio}
\newacronym{bler}{BLER}{block error rate}

\newacronym{kNN}{\textit{k}-NN}{\textit{k}-nearest neighbor algorithm}
\newacronym{SVM}{SVM}{support vector machines}
\newacronym{nn}{NN}{neural network}
\newacronym{NN}{NN}{neural network}
\newacronym{dnn}{DNN}{deep neural network}
\newacronym{RBF}{RBF}{radial basis function}
\newacronym{RMSE}{RMSE}{root mean squared error}
\newacronym{mse}{MSE}{mean squared error}
\newacronym{lmse}{LMSE}{linear mean square-error estimator}

\newacronym{R2}{$R^2$}{coefficient of determination}

\newacronym{KAUST}{KAUST}{King Abdullah University of Science and Technology}
\newacronym{GSA}{GSA}{Global mobile Suppliers Association}

\newacronym{VoD}{VoD}{video on demand}
\newacronym{HEVC}{HEVC}{High Efficiency of Video Coding}
\newacronym{DASH}{DASH}{Dynamic Adaptive Streaming over HTTP}

\newacronym{PUT}{PUT}{people using television}

\newacronym{ADTVS}{ADTVS}{Audience Driven live TV Scheduling}

\newacronym{arq}{ARQ}{automatic repeat request}

\newacronym{harq}{HARQ}{hybrid automatic repeat request}

\newacronym{sdp}{SDP}{semi-definite programming}

\newacronym{tcp}{TCP}{transmission control protocol}

\newacronym{e2e}{E2E}{end-to-end}

\newacronym{ran}{RAN}{radio access network}
\newacronym{cran}{CRAN}{cloud radio access network}
\newacronym{udcran}{UD-CRAN}{ultra-dense CRAN}
\newacronym{dran}{DRAN}{distributed radio access network}
\newacronym{hcran}{H-CRAN}{hybrid cloud radio access network}
\newacronym{hetnet}{HetNet}{heterogeneous network}
\newacronym{vcran}{V-CRAN}{virtualized CRAN}
\newacronym{ecran}{E-CRAN}{edge-CRAN}
\newacronym{hvcran}{H-VCRAN}{hybrid-virtualized CRAN}

\newacronym{bbu}{BBU}{baseband processing unit}
\newacronym{rrh}{RRH}{remote radio head}
\newacronym{ru}{RU}{radio unit}
\newacronym{rs}{RS}{remote site}
\newacronym{cs}{CS}{central site}

\newacronym{rru}{RRU}{radio resource unit}
\newacronym{rb}{RB}{resource block}
\newacronym{hpn}{HPN}{high-power node}
\newacronym{lpn}{LPN}{low-power node}
\newacronym{mabs}{MaBS}{macro basestation}
\newacronym{ue}{UE}{user equipment}

\newacronym{comp}{CoMP}{coordinated multi-point}
\newacronym{ranaas}{RANaaS}{RAN-as-a-Service}

\newacronym{rof}{RoF}{radio over fiber}
\newacronym{wdm}{WDM}{Wavelength Division Multiplexing}
\newacronym{dls}{DLS}{distributed large scale}

\newacronym{qos}{QoS}{quality of service}
\newacronym{qoe}{QoE}{quality of experience}
\newacronym{qee}{QEE}{quality of energy-efficiency}

\newacronym{gg}{GG}{group-to-group}
\newacronym{ht}{HT}{hyper-transceiver}

\newacronym{fh}{FH}{fronthaul}
\newacronym{dl}{DL}{downlink}
\newacronym{ul}{UL}{uplink}

\newacronym{cp}{CP}{Cell-Processing}
\newacronym{up}{UP}{User-Processing}

\newacronym{co}{CO}{center office}

\newacronym{du}{DU}{digital unit}
\newacronym{lc}{LC}{Line-Card}

\newacronym{onu}{ONU}{optical network unit}
\newacronym{olt}{OLT}{optical line terminal}
\newacronym{osw}{OSW}{optical switch}

\newacronym{es}{ES}{ethernet switch}

\newacronym{ppp}{PPP}{Poisson point process}

\newacronym{mppp}{MPPP}{marked Poisson point process}

\newacronym{sinr}{SINR}{signal to noise and interference ratio}
\newacronym{sir}{SIR}{signal to interference ratio}

\newacronym{mbs}{MBS}{macro basestation}
\newacronym{ap}{AP}{access point}
\newacronym{fap}{FAP}{femto-cell access point}
\newacronym{sap}{SAP}{small-cell access point}
\newacronym{iot}{IoT}{Internet of Things}
\newacronym{ti}{TI}{Tactile Internet}
\newacronym{lsm}{LSM}{linear scalarizing method}

\newacronym{lp}{LP}{Low-Priority}
\newacronym{hp}{HP}{High-Priority}
\newacronym{lpu}{LPU}{Low-Priority user}
\newacronym{hpu}{HPU}{High-Priority user}
\newacronym{lps}{LPS}{Low-Priority system}
\newacronym{hps}{HPS}{High-Priority system}

\newacronym{ttm}{TTM}{time to market}
\newacronym{udn}{UDN}{ultra-dense network}

\newacronym{capex}{CAPEX}{capital expenditure}
\newacronym{opex}{OPEX}{operational expenditure}

\newacronym{cpri}{CPRI}{common public radio interface}
\newacronym{otn}{OTN}{optical transport network}
\newacronym{pon}{PON}{passive optical network}
\newacronym{twdm}{TWDM}{time and wavelength division multiplexing}

\newacronym{ec}{EC}{Edge-Cloud}
\newacronym{cc}{CC}{Central-Cloud}

\newacronym{mmw}{m-Wave}{Milli-Meter wave}

\newacronym{gops}{GOPS}{giga operation per second}
\newacronym{mops}{MOPS}{mega operation per second}

\newacronym{ip}{IP}{internet protocol}
\newacronym{rlc}{RLC}{radio link control}
\newacronym{pdcp}{PDCP}{packet data convergence protocol}

\newacronym{mno}{MNO}{mobile network operator}
\newacronym{prb}{PRB}{physical resource block}
\newacronym{mi}{MI}{modulation index}

\newacronym{wifi}{WiFi}{wireless local area network}
\newacronym{cpu}{CPU}{central processing unit}
\newacronym{vcpu}{VCPU}{virtual CPU}
\newacronym{vm}{VM}{virtual machine}

\newacronym{urs}{UrS}{user requested service}

\newacronym{rsf}{RSF}{radio sub-frame}
\newacronym{siso}{SISO}{single-input single-output}

\newacronym{mec}{MEC}{mobile edge computing}
\newacronym{co2}{CO$_{2}$}{carbo dioxide}

\newacronym{ar}{AR}{augmented reality}
\newacronym{vr}{VR}{virtual reality}
\newacronym{cfp}{CFP}{communication function processing}
\newacronym{ptp}{PTP}{precision time protocol}

\newacronym{voip}{VoIP}{voice over Internet protocol}

\newacronym{da}{DA}{data analytics}

\newacronym{kpi}{KPI}{key performance indicator}

\newacronym{fsmc}{FSMC}{finite state markov chain}
\newacronym{ml}{ML}{machine learning}
\newacronym{dgd}{DGD}{distributed gradient descent}

\newacronym{5g}{5G}{fifth generation of mobile communication systems}
\newacronym{gnbcu}{gNB-CU}{gNB central unit}
\newacronym{gnbdu}{gNB-DU}{gNB distributed unit}
\newacronym{ecpri}{eCPRI}{common public radio interface}
\newacronym{fl}{FL}{federated learning}
\newacronym{rsrq}{RSRQ}{reference signal received quality}
\newacronym{rsrp}{RSRP}{reference signal received power}
\newacronym{urllc}{URLLC}{ultra-reliable low latency communications}
\newabbreviation{embb}{eMBB}{enhanced mobile broadband}
\newabbreviation{mmtc}{mMTC}{massive machine type communication}

\newacronym{mae}{MAE}{modified autoencoder}
\newacronym{mtc}{MTC}{machine type communication}
\newacronym{pca}{PCA}{principal component analysis}

\newacronym{cps}{CPS}{cyber-physical system}
\newacronym{gnb}{gNB}{gNodeB}
\newacronym{ref}{REF}{reliability enhancement feature}
\newacronym{nfo}{NFO}{network level feature orchestrator}
\newacronym{dc}{DC}{data center}
\newacronym{nf}{NF}{network function}
\newacronym{vnf}{VNF}{virtual network function}
\newacronym{nfv}{NFV}{network functions virtualization}
\newacronym{nssmf}{NSSMF}{network slice subnet management function}
\newacronym{nsmf}{NSMF}{network slice management function}
\newacronym{ai}{AI}{artificial intelligence}
\newacronym{rl}{RL}{reinforcement learning}
\newacronym{drl}{DRL}{deep reinforcement learning}
\newacronym{ddpg}{DDPG}{deep deterministic policy gradient}
\newacronym{dqn}{DQN}{deep Q-networks}

\newacronym{a2c}{A2C}{advantage actor-critic}

\newacronym{td3}{TD3}{twin delayed deep deterministic policy gradient algorithm}
\newacronym{sgd}{SGD}{stochastic gradient descent}
\newacronym{um}{UM}{unacknowledged mode}
\newacronym{am}{AM}{acknowledged mode}
\newacronym{mdp}{MDP}{Markov decision process}
\newacronym{tti}{TTI}{transmission time interval}

\newabbreviation{bsac}{BSAC}{branching \acrlong{sac}}
\newabbreviation{sac}{SAC}{soft actor-critic}
\newabbreviation{inf}{InF}{indoor factory}
\newabbreviation{pmf}{PMF}{probability mass function}
\newabbreviation{cdf}{CDF}{cumulative distribution function}

\newabbreviation{ofdm}{OFDM}{orthogonal frequency-division multiplexing}

\newabbreviation[description={\glsxtrshort{inf}-with dense clutter and high base station height}]{inf-dh}{InF-DH}{indoor factory with dense clutter and high base station height}

\newabbreviation{los}{LOS}{line-of-sight}
\newabbreviation[description={non-\glsxtrshort{los}}]{nlos}{NLOS}{non-LOS}

\newboolean{appendixversion}
\setboolean{appendixversion}{true} 

\begin{document}
\begin{NoHyper}
\bstctlcite{IEEEexample:BSTcontrol}	
\IEEEoverridecommandlockouts

\title{BSAC-CoEx: Coexistence of URLLC and Distributed Learning Services via Device Selection}
\author{
\IEEEauthorblockN{Milad Ganjalizadeh, Hossein S. Ghadikolaei, Deniz Gündüz, and Marina Petrova
\ifthenelse{\boolean{appendixversion}}{
\thanks{This work has been submitted to the IEEE for possible publication. Copyright may be transferred without notice, after which this version may no longer be accessible.}
}{
\thanks{Milad Ganjalizadeh and Hossein S. Ghadikolaei are with Ericsson Research, Ericsson AB, 164 40 Stockholm, Sweden (email: \{milad.ganjalizadeh; hossein.shokri.ghadikolaei\}@ericsson.com).}
\thanks{Deniz Gündüz is with the Department of Electrical and Electronic Engineering, Imperial College London, SW7 2AZ London, UK (email: d.gunduz@imperial.ac.uk).}
\thanks{Marina Petrova is with School of EECS, KTH Royal Institute of Technology, 114 28 Stockholm, Sweden and RWTH Aachen University, 52062 Aachen, Germany (email: petrovam@kth.se).}
}
}\\
\vspace{-5mm}
}
\markboth{}%
{Ganjalizadeh \MakeLowercase{\textit{et al.}}: BSAC-C\MakeLowercase{o}E\MakeLowercase{x}: Coexistence of URLLC and Distributed Learning Services via Device Selection}

\maketitle
	
\begin{abstract}
Recent advances in distributed intelligence have driven impressive progress across a diverse range of applications, from industrial automation to autonomous transportation. Nevertheless, deploying distributed learning services over wireless networks poses numerous challenges. These arise from inherent uncertainties in wireless environments (e.g., random channel fluctuations), limited resources (e.g., bandwidth and transmit power), and the presence of coexisting services on the network. In this paper, we investigate a mixed service scenario wherein high-priority ultra-reliable low latency communication (URLLC) and low-priority distributed learning services run concurrently over a network. Utilizing device selection, we aim to minimize the convergence time of distributed learning while simultaneously fulfilling the requirements of the URLLC service. We formulate this problem as a Markov decision process and address it via BSAC-CoEx, a framework based on the branching soft actor-critic (BSAC) algorithm that determines each device's participation decision through distinct branches in the actor's neural network. We evaluate our solution with a realistic simulator that is compliant with 3GPP standards for factory automation use cases. Our simulation results confirm that our solution can significantly decrease the training delays of the distributed learning service while keeping the URLLC availability above its required threshold and close to the scenario where URLLC solely consumes all wireless resources.
\end{abstract}
	
\section{Introduction}\label{sec:intro}
\IEEEPARstart{T}{he} envisioned 6G networks seek to go beyond mere connectivity by embedding intelligence into all aspects of communication. This evolution paves the way for a cyber-physical ecosystem that seamlessly connects devices, \gls{ai}, and humans~\cite{surveyEdgeShi,denizMag}.
In parallel, a growing wave of \gls{ai} services has demonstrated remarkable performance gains across diverse use cases, from \gls{embb}~\cite{ganjCascaded} to industrial control~\cite{dAIFaultDetectFactory} and self-driving vehicles~\cite{vehicularFlTsnm}. However, these gains hinge on large-scale model training, a process often hindered by data privacy concerns, limited wireless resources, and the widespread distribution of devices possessing data. 

To mitigate these challenges, privacy-preserving distributed learning has emerged as a promising solution, allowing edge devices to train \gls{ai} models collaboratively without sharing raw local data. Typically, multiple devices upload local updates (e.g., local models in \gls{fl}~\cite{fl} or local gradients in \gls{dgd}~\cite{duttaKsync}) using the \gls{ul} channel to a central node (or a set of nodes). The central node maintains the global parameters and orchestrates each iteration of distributed training. Once it updates the global model, it transmits the new parameters to devices via the \gls{dl} channel, initiating the next iteration.

The wall-clock convergence time of a distributed learning algorithm depends on i) the number of required training iterations to reach a certain optimality gap, and ii) the delay of each training iteration (which includes the \gls{ul}/\gls{dl} transmission of local updates/global model parameters, and computation at devices and the central node). The former depends on various factors, such as the model size, data quality, selected learning algorithm and its hyperparameters~\cite{flHyperparamDevSelTsnm, chen2021Survey}.
Under ideal processing and communication assumptions, increasing the number of participating devices at each iteration often leads to a better convergence rate, resulting in a shorter convergence time.
However, when implemented across wireless devices, because of interference and limited available bandwidth, the training delay is impacted not only by the model parameters and the learning algorithm but also by the wireless channel resources between the central node and individual devices. In this case, while more participating devices may reduce the required number of iterations, it can extend the convergence time due to increased communication and computation delays~\cite{chen2021Survey}.

\Gls{urllc} is characterized by its stringent requirements, demanding delays as short as $500$\,$\mu$s, and availability up to $99.9999999$~\cite{3GPP22104}.
In~\cite{3GPP22104,3GPP22261}, \gls{3gpp} defines \emph{communication service availability} as the mean proportion of time during which the communication service satisfies the required \gls{qos} of the application it serves.
\Gls{urllc} services are central to mission-critical industrial applications (such as robotics and factory automation) where even minor delays can lead to costly outcomes~\cite{urllcApp}. 

In this paper, we investigate optimizing the coexistence of \gls{urllc} and distributed learning services from communication perspective.
This integration exemplifies the broader \emph{\gls{ai}-RAN} paradigm~\cite{aiRanAlliance}, which aims to enable \gls{ai} services to share infrastructure with conventional services, utilizing additional communication and computation resources.

\subsection{Motivations}
There is a rich literature on conventional mixed services between \gls{urllc} and \gls{embb} (e.g.,~\cite{urllcEmbbTransCoex,embbUrllcTsnm,embbUrllcDistAiTsnm}), \gls{urllc} and \gls{mmtc} (e.g.,~\cite{mimpUrllcMmtc}), or even all three (e.g.,~\cite{popovski20185gUrllcEmbbMmtc}).
However, the existing solutions may not be suitable to govern the coexistence of \gls{urllc} and distributed learning services for the following reasons:

\begin{itemize}
    \item There are two fundamental differences between distributed learning and other conventional communication services. First, the performance of these conventional services is characterized by the statistics of a communication metric (e.g., throughput for \gls{embb} and energy consumption for \gls{mmtc}). In contrast, distributed learning services are iterative and collaborative tasks aiming to solve an optimization problem. Besides, the potential statistical correlation of data from various devices enables distributed training to operate using only a subset of devices. 
    Second, it has a unique set of decision variables such as the model size, choice of algorithm, and selection of devices participating in the training task.

    \item \gls{urllc} traffic affects the convergence time of distributed training in unique ways due to its higher traffic priority and strict requirements. 
    For example, if \gls{urllc} devices have more intensive traffic or stricter availability requirements, low-priority services, including distributed learning, may face performance degradation due to fewer available \glspl{prb}. Alternatively, more participating devices in the training could increase the interference footprint and harm \gls{urllc} performance. A proper device selection strategy can alleviate such performance degradation. This is possible due to the unique properties of distributed learning applications and their robustness to occasional missing updates.
    The tighter the requirements of the coexisting service, the harder it is to design smooth coexistence.
\end{itemize}

Despite extensive research efforts to enhance the performance of \gls{urllc} and distributed learning over wireless networks as individual services, no studies to date have addressed the coexistence of these two services with both communication and learning aspects jointly considered. 
Nevertheless, numerous 6G applications leverage distributed learning to improve efficiency and system performance in areas such as industrial automation~\cite{dAIFaultDetectFactory} and autonomous driving~\cite{vehicularFlTsnm}. The successful adoption of these applications relies on analyzing and optimizing them alongside existing services, particularly \gls{urllc}, which poses the most demanding requirements. This serves as our primary motivation for this work.



\subsection{Contributions}
In this paper, we investigate an optimal coexistence of distributed learning and \gls{urllc} services. Our previous work \cite{ganjInterplay} examined the mutual impact of these two services under 5G-NR protocols. Despite employing strict priority scheduling, we observed that the distributed training traffic could significantly influence the \glspl{kpi} of \gls{urllc} and vice versa. 
In this work, we substantially extend \cite{ganjInterplay} by expanding the system model, adding a new decision variable, developing new solution algorithms, and providing new engineering insights. In particular, we adopt a soft synchronous distributed learning protocol in which a central node broadcasts the global model updates upon receiving local updates from a subset of available devices. We then optimize the interplay between the two services. 
In summary, our contributions are as follows:
\begin{itemize}
    \item We develop a model for the operational metrics of the \gls{urllc} service (i.e., communication service availability) and essential parameters that characterize distributed training workflow (i.e., training delay and convergence) and investigate the interplay between them. Since the system is resource-limited (in terms of bandwidth and transmission power) and \gls{urllc} requirements are strict, a subset of devices must be selected to perform each iteration of distributed training. Accordingly, we formulate an optimization problem that minimizes the average training latency to reach a given optimality gap while meeting \gls{urllc}'s strict availability requirements.
    
    \item We transform the formulated coexistence optimization problem into a \gls{mdp} and design an action masking technique to put a lower bound on the minimum number of devices required to participate in each iteration of distributed training. However, the solution may select a higher number of devices than this minimum to address the so-called \textit{straggling effect}, where the lack of updates from some devices may halt the entire training.
    
    \item To deal with the unknown dynamics of our mobile network, we propose a data-driven approach that optimizes the device selection policy via a state-of-the-art off-policy \gls{drl} algorithm, called BSAC-CoEx, a \gls{bsac} framework designed for device selection in our coexistence scenario. In \gls{bsac}-CoEx, the selection policy of each device is distributed across distinct neurons, resulting in a linear increase in the \gls{bsac} output size with the number of devices.
    
    \item We evaluate our framework using a \gls{3gpp}-compliant 5G simulator in a factory automation use case. The results reveal that our solution not only preserves \gls{urllc} availability nearly on par with standalone \gls{urllc} service, but also reduces median training delays by more than $20$\%, compared to the slicing setting with service-specific resource partitions.
    Our results provide valuable insights into the ongoing standardization activities of distributed learning~\cite{3GPP22874AiModel,3GPP22876AiModelP2}.
\end{itemize}

The rest of this paper is organized as follows. We provide the necessary preliminaries on distributed learning in Section~\ref{sec:Background} and define our system model in Section~\ref{sec:systemModel}. The performance metrics and problem formulation are described in Section~\ref{sec:PM}. Section~\ref{sec:solution} presents the proposed \gls{mdp} modeling and \gls{bsac}-CoEx framework.
We describe our simulation methodology in Section~\ref{sec:simulation}, and discuss the results in Section~\ref{sec:preformance}. Finally, Section~\ref{sec:conculsions} concludes the paper.

\textit{Notations:} 
Normal font $x$ or $X$, bold font $\bm{x}$, bold font $\bm{X}$, and uppercase calligraphic font $\mathcal{X}$ denote scalars, vectors, matrices and sets, respectively.
We denote by $[X]$ the set $\{1,2,\ldots,X\}$, by $\left[\bm{x}\right]_{i}$ element $i$ of vector $\bm{x}$, by $\left[\bm{X}\right]_{i,j}$ the element $ij$ of matrix $\bm{X}$, and by $|\calX|$ the cardinality of set $\calX$. We define $\mathds{1}\{\bm{x}\}$ as the element-wise indicator function returning $\bm{y}$, where $[\bm{y}]_i$ takes 1 when condition $[\bm{x}]_i$ holds, and 0, otherwise. The curled inequality ($\succeq$ or $\succ$) represents element-wise inequality. We use $\bm{1}$ and $\bm{0}$ to denote all-one and all-zero vectors, respectively, and $\N$ to denote the set of natural numbers. Table~\ref{tab:notations} 
summarizes our main notations.
\section{Distributed Learning Over Wireless Networks}\label{sec:Background}
We consider the problem of minimizing a sum of smooth functions $\{f_{i}: \R^d \mapsto \R\}_{i \in [N]}$, with corresponding gradients $\{\nabla f_{i}: \R^d \mapsto \R^d\}_{i \in [N]}$:
\begin{equation}\label{eq: MainOptimProblem}
\bw^\star \coloneqq \min_{\bw\in\R^d} f(\bw) = \min_{\bw\in\R^d} \frac{1}{N}\sum\nolimits_{i \in [N]}f_i(\bw),
\end{equation}
where $f$ is the global loss function, each $f_i$ represents a local loss function, and $N$ is the number of distributed devices. In practice, we may use distributed algorithms to solve~\eqref{eq: MainOptimProblem} to expedite the computations or to preserve the privacy of local datasets~\cite{Bottou2018SIAM}. At the $k$th iteration, a subset of workers compute their local gradients and share them with a central node, which maintains the global model. The central node updates and broadcasts $\bw_{k+1}$ back to the workers.\footnote{We have a similar set of trade-offs and solutions for non-smooth functions, where we cannot define gradients. The major difference compared to this paper is that instead of updating based on gradients, we may need to update based on its generalizations, like subgradients~\cite{scaman2018optimal}.} \gls{fl} is another method in which the workers share their updated local models instead of the gradients. Subsequently, the central node takes a global average over them. The communication overhead is almost the same as uploading gradients~\cite{li2019federated}. 

On the one hand, most of these \gls{ul} messages (gradients or local models) may be redundant since they can be retrieved from past messages or messages from other devices~\cite{ghadikolaei2021lena}. On the other hand, the central node must wait for all the participating devices to send their updates, which results in considerable idle time for the central node and delays for faster devices waiting for stragglers to complete their updates.
To tackle the straggling issue, in soft synchronous approaches, the central node only waits for a subset of participating devices, say $n$ out of all $N$ devices, to update the global model at each iteration~\cite{duttaKsync}. Reference~\cite{ji2020dynamic} proposed an algorithm to adjust $n$ at each iteration.
However, vanilla $n$-sync-based approaches increase the load on the underlying communication system by demanding all devices to upload their typically large model parameters and starting the update process at the central node only after receiving the first $n$ messages.
Forcing some of the devices to remain silent would i) reduce \gls{ul} interference to other users, ii) improve latency, and iii) increase throughput.
References~\cite{fl,ghadikolaei2021lena,chen2018lag} proposed various techniques to eliminate unnecessary uploads.

Running distributed learning over wireless networks presents additional challenges; not only due to the scarcity of resources (e.g., bandwidth and transmit power), but also because of the wireless propagation dynamics---affected by factors such as noise, interference, and fading~\cite{chen2021Survey}. To address these challenges, several recent studies have focused on resource management and, more explicitly, device selection techniques to improve the performance of distributed training in terms of training loss or convergence time~\cite{chenJointLearningCommFL,dinhFEDL,energyDelayEarlyAccessTmcTd3,deviceSelMab,energyDelayTiiDdpg,optFlIIoT,denizUpdateAware}. For example, in~\cite{chenJointLearningCommFL}, the authors evaluated the impact of resource management, device selection, and transmit power on \gls{fl} convergence, and optimized these parameters to reduce \gls{fl} training loss. Reference~\cite{dinhFEDL} proposed an \gls{fl} algorithm that can handle heterogeneous user data assuming strong convex and smooth loss functions. The resource allocation was then optimized to improve the convergence of the proposed algorithm. Unlike previous model-based approaches, references~\cite{energyDelayEarlyAccessTmcTd3,deviceSelMab,energyDelayTiiDdpg,optFlIIoT} utilized \gls{drl} to perform device selection. 
A device selection policy, jointly based on computation speed and channel condition, is proposed in~\cite{denizSpawc}.
Global model delivery to the devices over a noisy \gls{dl} channel is studied in~\cite{denizAmiri}.
In~\cite{denizUpdateAware}, it is shown that distributed training performance can be improved if the device selection algorithm jointly considers the updates from the devices with their channel conditions.
Reference~\cite{flFairSelTsnm} introduces a device selection and scheduling framework for the fair selection of edge devices in clustered federated multitask learning, ensuring fair participation opportunities during training. This approach enhances convergence speed and mitigates biases in the resulting global models.
Nevertheless, none of these works address scenarios involving mixed services, where wireless network characteristics and limitations, along with demands on the higher-priority service, determine distributed learning performance.

\section{System Model}\label{sec:systemModel}
\subsection{Network Model}\label{sec:networkModel}
In the context of industrial automation, which is one of major use cases of \gls{urllc} services~\cite{urllcApp}, we consider a scenario where a set of $\mathcal{G}{\coloneqq}[G]$ \glspl{gnb}, each consisting of one cell, serve a set of $\mathcal{U}{\coloneqq}[U]$ \gls{urllc} devices, executing different functions that enable automated production. The communication system should timely and reliably deliver i) computed or emergency control commands to the actuators, and ii) monitoring data to \glspl{gnb}.
To simplify our analysis, we assume that the devices participating in the learning process are distinct from the \gls{urllc} devices, and for distinction, we will call the former \gls{ai} devices. We consider a set of $\mathcal{N} {\coloneqq} [N]$ \gls{ai} devices, orchestrated by a single parameter server (i.e., central node) to carry out a single distributed learning task. 
We assume that the central node waits until receiving the local model update from $n$ out of these $N$ \gls{ai} devices at each iteration. To tackle the straggler effect, the central node might request an update from $\calN_{m,k} \subseteq \calN$ of the \gls{ai} devices at iteration $k$, where $|\mathcal{N}_{m,k}| {=} m_k({\geq} n)$ (i.e., the central node might request $m_k{-}n$ extra backup devices to mitigate the straggler problem). 


In the context of 5G and beyond, two approaches are proposed to manage the coexistence of two services with different priorities. The first approach is to leverage the existing standardized protocols in 5G-NR for \gls{qos} handling. 
The second approach involves creating separate slices for \gls{urllc} and distributed learning services, thus resulting in complete resource isolation.
We employ the former approach. In this scenario, each connected device is assigned one or more \gls{qos} flows and data radio bearers. The \gls{qos} flows are determined based on the service's \gls{qos} requirements and configured in the core network. For instance, in our specific case, traffic from or to \gls{urllc} devices is given a high-priority \gls{qos} flow to ensure minimal latency. These \gls{qos} flows are then mapped to data radio bearers in the \gls{ran}. Both the \gls{gnb} and devices have an associated \gls{rlc} buffer for each data radio bearer, with strict priority scheduling applied in our case, as described in~\cite{dahlman5GNr}.

\subsection{Distributed Learning Process}\label{sec:distAlg}
We consider a network of $N$ \gls{ai} devices that cooperatively solve a distributed learning problem. Assuming that $\calN_{n,k} \subseteq \calN_{m,k}$ is a subset of size $n$, whose updates the central node receives first at iteration $k$, then iteration $k$ of an abstract distributed algorithm reads:
\vspace{-2mm}
\begin{subequations}\label{eq:dAI}
\begin{align}
    &\bw_{k+1}{=}A{\left(\bc_{i,k}, \bw_k\right)},\quad \mbox{for} \quad \forall i \in \calN_{n,k}, \label{eq:globalUp}\\
    & \bc_{i,k}{=}C_i{\left(\bw_{k}\right)},\quad \mbox{for} \quad \forall i \in \calN_{m,k}, \label{eq:localUp}
\end{align}
\end{subequations}
where function $A$ represents an algorithm update of the decision variable $\bw_k$, function $C_i$ picks out the relevant information, $\bc_{i,k}$, that device $i$ uploads to the server for algorithm execution. This general algorithmic framework accommodates a variety of  \gls{ml} algorithms, including \gls{fl} and \gls{dgd}, with or without data compression. For instance, when $C_i$ yields a stochastic gradient, say $\widehat{\nabla} f_i(\bw_{k})$, and $A=\bw_k {-} \eta \sum_i \widehat{\nabla} f_i(\bw_{k})/n$ for some positive step size $\eta$, we obtain $n$-sync and synchronous \gls{dgd} for $n({<} N)$ and $n({=} N)$, respectively~\cite{duttaKsync}. When $C_i$ returns the updated local model parameters of the $i$th \gls{ai} device and $A$ performs an averaging step over a subset of $n ({\leq} N)$ \gls{ai} devices, we recover federated averaging~\cite{fl} (either $n$-sync or synchronous).

\subsection{Channel Model}\label{sec:systemModelCm}
To model the channel, we consider a \gls{mimo} system in which we leverage the 3D spatial channel model from \gls{3gpp} in~\cite{3GPP38901}. In this model, channels are characterized via clustering the multipath components, arriving at antenna arrays, in delay and double-directional angle (i.e., the zenith and azimuth of the \glspl{aoa} at the receiver and \glspl{aod} at the transmitter). 
\ifthenelse{\boolean{appendixversion}}{
We have moved the details of the channel model to Appendix~\ref{app:channelModel}.
}{
We have moved the details of the channel model to \cite[Appendix B]{extended}.
}
Notice that our problem formulation (Section~\ref{sec:problemFor}) and solution approach (Section~\ref{sec:solution}) are general and not limited to this channel model. 

For clarity and convenience, the symbols and notations used throughout this paper are summarized in Table~\ref{tab:notations}.
\begin{table}[t]
\footnotesize
\centering
\caption{Main Symbols and Notations}
\label{tab:notations}
\renewcommand{\arraystretch}{1.1}
\begin{tabular}{p{0.07\textwidth} || p{0.35\textwidth}}
\hlineB
\textbf{Symbol} & \textbf{Description} \\
\hline

$\mathcal{N}, \mathcal{U}, \mathcal{G}$ 
      & Sets of AI devices, URLLC devices, and gNBs.\\

$\bm{\pi}_k^\mathrm{u},\, m_k$
  & Device-selection vector at iteration $k$ and the number of selected devices.\\

$n$ 
  & Minimum required number of local updates per iteration ($n$-sync), $m_k \ge n$.\\

$d_k^{\mathrm{AI}}$ 
      & Distributed training delay in iteration $k$.\\

$\bw_k,\, \bw^\star$
  & Global model at iteration $k$, and optimal global model.\\

$K_{\min},\, \epsilon$
  & Number of global iterations for convergence, and tolerance parameter defining $\epsilon$-optimality.\\

$\alpha_i^\Gamma$
  & URLLC device $i$’s availability in $\Gamma$ (${\in} \{\mathrm{UL},\mathrm{DL}\}$).\vspace{0.2mm}\\

$\alpha_i^\mathrm{req}$
  & URLLC device $i$’s availability requirement.\\

$\gamma$ 
  & Tolerable violation probability in $\Pr\{\alpha_i^\Gamma \le \alpha_i^\mathrm{req}\} \le \gamma$.\\

$T_{\mathrm{sv}}$
  & Survival time for URLLC application-layer continuity.\\

$\bm{s}_k,\, \bm{a}_k, r_{k}$
  & State, action, and reward of iteration $k$, respectively.\\

$\lambda,\, \psi$
  & Discount factor and temperature parameter in SAC.\\

\ifthenelse{\boolean{appendixversion}}{
$\bm{H}_{x,y}(\tau;t)$
  & MIMO channel impulse response between device $x$ and gNB $y$.\\
  
$N_\mathrm{g}, N_\mathrm{d}$ 
& Number of gNB antennas and device antennas.\\

$N_\mathrm{c}, N_\mathrm{s}$ 
& Number of clusters and rays in the spatial channel model.\\

$\beta^{x,y}$ 
& Large-scale fading for device $x$ to gNB $y$.\\

$\mathrm{Pr}_{\mathrm{LOS}}$ 
& Probability a link is in LOS.\\

$PL_{\mathrm{LOS}}$ 
& Pathloss under LOS conditions.\\

$PL_{\mathrm{NLOS}}$ 
& Pathloss under NLOS condition.\\

$d_{x,y}^{\mathrm{2D}},\, d_{x,y}^{\mathrm{3D}}$ 
& 2D and 3D distance between device $x$ and gNB $y$.\vspace{0.3mm}\\
\hlineB
}{
$\bm{H}_{x,y}(\tau;t)$
  & MIMO channel impulse response (3GPP model) between device $x$ and gNB $y$.\\
\hlineB
}

\end{tabular}
\end{table}


\section{Performance Metrics and Problem Formulation}\label{sec:PM}
\subsection{URLLC KPI: Communication Service Availability}\label{sec:urllcKpi}
The convergence of operational, information, and communication technologies through 5G is paving the way for connected industries. However, this integration poses a significant challenge of ensuring that the operational requirements are met during the 5G system's operating phase~\cite{5gAcia}.
One well-accepted metric in the operational technology domain is \textit{availability}. Hence, \gls{3gpp}, as the primary standardization consortium for mobile telecommunications, has attempted to specify the requirements for communication service availability from the application layer perspective in~\cite{3GPP22104,3GPP22261}.
The main difference between the network performance and the observed performance on the application layer is driven by a system parameter called \textit{survival time}, $T_{\mathrm{sv}}$. Survival time is the period of time for which the application layer can continue to function without receiving an expected packet~\cite{ganjPimrcTranslation}.
We denote the network layer state by a Bernoulli random variable $X_i^{\Gamma}{\left(t\right)}$, where $\Gamma{\in}$\{\gls{ul}, \gls{dl}\}, and $X_i^{\Gamma}{\left(t\right)}$ for the $i$th \gls{urllc} device is zero (i.e., link outage) if the last packet is not received at the communication interface within a specified delay bound, because either it could not be decoded at the lower layers or faced excessive retransmission and/or queuing delays.
Assuming negligible application recovery time, we then define the per-device application layer state variable as   
\begin{equation}\label{eq:appState}
Y_i^{\Gamma}{\left(t\right)}{\coloneqq}
    \begin{cases}
      0, & \mathrm{if} \int_{t-T_{\mathrm{sv}}}^t X_i^{\Gamma}\!{\left(\tau\right)} \diff \tau = 0\enspace\mathrm{(failure)},\\
      {1,} &  \mathrm{otherwise}\enspace\mathrm{(operational)}. \\
    \end{cases}
\end{equation}
Thus, we can define the long-term communication service availability for the $i$th \gls{urllc} device in $\Gamma$ direction as~\cite{ganjTIIOrch}
\begin{equation}\label{eq:availInf}
    \alpha_i^{\Gamma}{\coloneqq} \lim_{t\to \infty}\Pr\left\{Y_i^{\Gamma}\!{\left(t\right)} {=} 1\right\} {=} \lim_{T\to\infty}\frac{1}{T}\int_{0}^T Y_i^{\Gamma}\!{\left(t\right)} \diff t.
\end{equation}
The availability in $\Gamma$ direction can be estimated over a short time period using
\begin{equation}\label{eq:availEst}
    \hat{\alpha}_i^{\Gamma}{\left(\Delta t_k\right)} {\coloneqq} \frac{1}{\Delta t_k}\int_{t_k}^{t_k+\Delta t_k} Y_i^{\Gamma}\!{\left(t\right)} \diff t.
\end{equation}
In \gls{urllc}, the requirement is often defined in the form of~\cite{popovskiUrllc}
\begin{equation}\label{eq:availReq}
    \Pr\left\{\alpha_i^{\Gamma} \leq \alpha_i^{\mathrm{req}}\right\}\leq\gamma, \enspace\forall i\in \calU,\
\end{equation}
where $\alpha_i^{\mathrm{req}}$ is the communication service availability requirement of the use case that \gls{urllc} device $i$ belongs to, and $\gamma$ is the maximum tolerable violation probability, which is determined by the sensitivity of this use case to $\alpha_i^{\mathrm{req}}$. We follow~\cite{3GPP22104} in assuming that the requirement for \gls{ul} and \gls{dl} availability is the same for a given use case.

\subsection{Distributed Learning KPI: Convergence Time}\label{sec:dAIKpi}

The convergence time of distributed learning algorithms is bounded by the communication and processing latency~\cite{chen2021Survey}. Let us denote the device selection at iteration $k$ by an indicator vector $\bm{\pi}_k^{\mathrm{u}}{=}\left[\left[\bm{\pi}_{k}^{\mathrm{u}}\right]_1, \left[\bm{\pi}_{k}^{\mathrm{u}}\right]_2, \ldots, \left[\bm{\pi}_{k}^{\mathrm{u}}\right]_N\right]$, where $\left[\bm{\pi}_{k}^{\mathrm{u}}\right]_i {\in} \{0,1\}, \forall i{\in} \calN$. Assuming that the central node requests the subset $\calN_{m,k}$ (i.e., $\bm{1}^T \bm{\pi}_k^{\mathrm{u}} {=}m_k$) to participate in the training while it only waits for $n$ local updates, the training delay at the central node for the $k$th iteration, $d_k^{\mathrm{AI}}$, can be derived as
\begin{equation}\label{eq:aiTrainingDelay}
d_k^{\mathrm{AI}}{\left(\bm{\Pi}_k^{\mathrm{u}}\right)} {\coloneqq} \min\!{\left\{\!\min_{\substack{\calN_{n,k}\subseteq\calN_{m,k},\\|\calN_{n,k}|{=}n}}{\!\!\left(\max_{i\in\calN_{n,k}}{\!\!\left(d_{i,k}^{\mathrm{D}}{+}d_{i,k}^{\mathrm{pr}}{+}d_{i,k}^{\mathrm{U}}\right)}\right)}{+}d_k^{\mathrm{pr}}, T^{\max}\!\right\}},
\end{equation}
where $\bm{\Pi}_k^{\mathrm{u}}{\coloneq} \left[\bm{\pi}_1^{\mathrm{u}}, \ldots, \bm{\pi}_k^{\mathrm{u}}\right]$ is the device selection matrix, $d_{i,k}^{\mathrm{D}}$, $d_{i,k}^{\mathrm{pr}}$, and $d_{i,k}^{\mathrm{U}}$ denote the latency associated with the transmission of the global model from the central node to \gls{ue}, local training (as in \eqref{eq:localUp}), and transmission of local parameters from the \gls{ue} to the central node for the $k$th iteration of device $i$, respectively. Note that both $d_{i,k}^{\mathrm{D}}$ and $d_{i,k}^{\mathrm{U}}$ encapsulate the transmission processing, payload transmission, occurred retransmissions, and queuing delay.
Thus, both are inherently functions of $\bm{\Pi}_k^{\mathrm{u}}$. Additionally, $d_k^{\mathrm{pr}}$ represents the processing delay in executing \eqref{eq:globalUp} on the central node for the $k$th iteration. Consequently, in \eqref{eq:aiTrainingDelay}, for each subset of $\calN_{n,k}$ with a cardinality of $n$, the maximum aggregated communication and processing delay is calculated among devices. To determine $d_k^{\mathrm{AI}}{(\cdot)}$ among these subsets, we select the subset with the lowest delay, subject to a maximum permissible delay of $T^{\mathrm{max}}$ for every iteration. In Section~\ref{sec:solution}, we develop a computational method to calculate all these delay components. \figurename\,\ref{fig:dAiWorkflow} illustrates the training delay in $n$-sync distributed training.

\begin{figure}[t]
	\centering
	\includegraphics[width=.99\columnwidth,keepaspectratio]{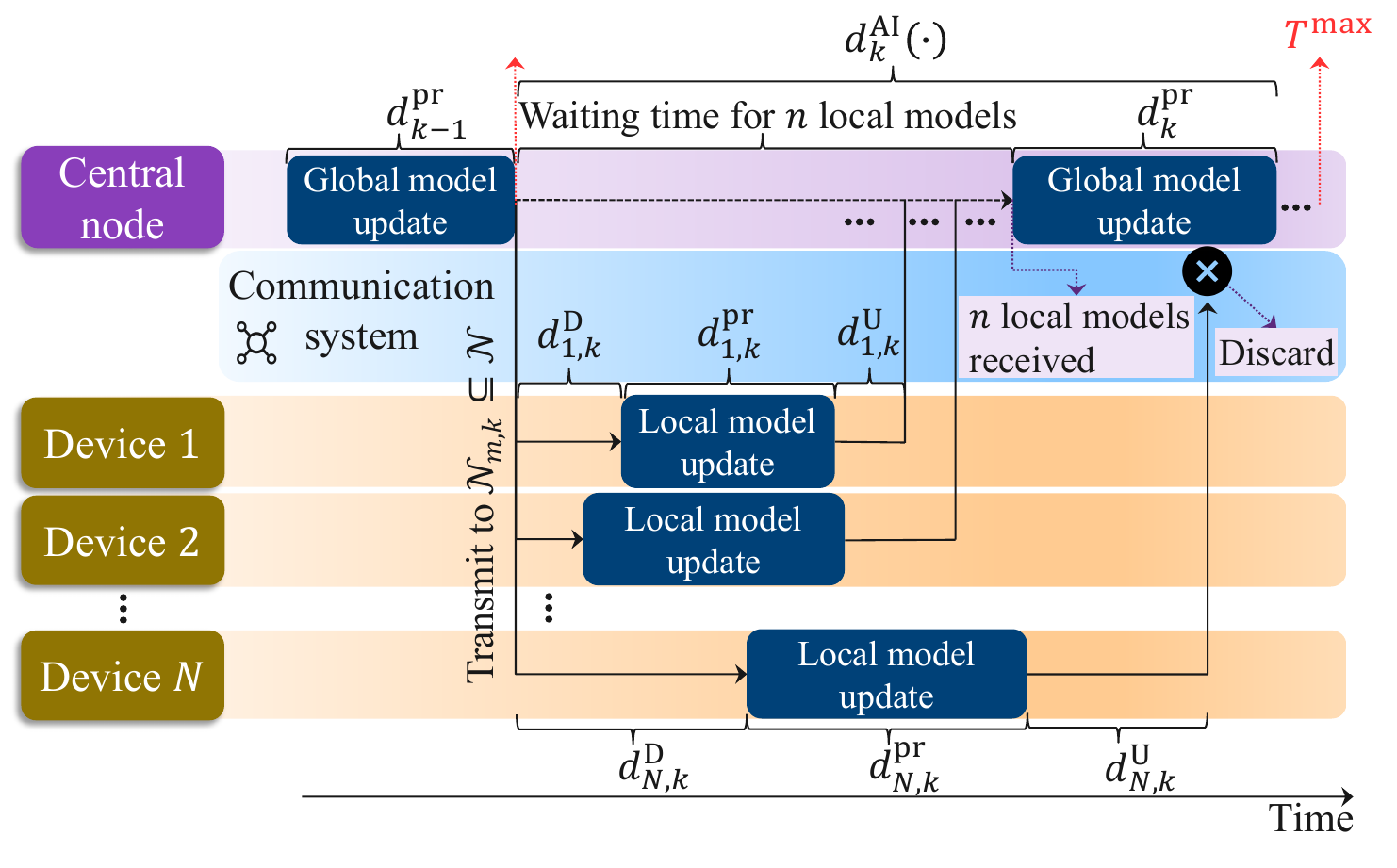}
	\caption{The illustration of training delay in distributed AI workflow.}
	\label{fig:dAiWorkflow}
	\vspace{-6mm}
\end{figure}

We denote by $K_{\min}$ the iteration number after which the distributed learning algorithm converges. To characterize $K_{\min}$, we define ``critical points" as the set of points where the norm of their derivative is 0. The local minima and maxima are a subset of these points~\cite{nonconvexOptBehrouz}. Note that a critical point may not necessarily be an optimal point, e.g., a saddle point.
To identify these critical points, we apply the first-order necessary condition by searching for the points at which the gradients vanish.
In practice, for computational efficiency, we assume that the global model converges when the approximate first-order necessary condition is met, and terminate the algorithm accordingly. Hence, we can formally define $K_{\min}$ as
\begin{equation}\label{eq:Kmin}
K_{\min} \coloneq \min\left\{k\middle|\enspace \mathbb{E}{\Bigl[{{\bigl\lVert\widehat{\nabla}f{(\bw_k)}\bigr\rVert}_2}\Bigr]}\leq\epsilon, k\in\mathbb{N}\right\},
\end{equation}
where $\epsilon$ is a small positive constant.
The $K_{\min}$ value varies depending on the specific problem setting and learning properties. We can use either numerical or theoretical approaches to determine $K_{\min}$.
In our experiments, we rely on numerical approximations to obtain $K_{\min}$.
\ifthenelse{\boolean{appendixversion}}{
  Nevertheless, as demonstrated in Appendix~\ref{app:a}, we can also theoretically approximate $K_{\min}$ for \gls{dgd} and \gls{fl} under smoothness, strong/non-convexity, and a few other technical assumptions related to the objective functions. Examples shown in Appendix~\ref{app:a} indicate that the value of $K_{\min}$ decreases as the number of devices participating in the global update (i.e., $n$) increases.
}{
  Nevertheless, as demonstrated in \cite[Appendix A]{extended}, we can also theoretically approximate $K_{\min}$ for \gls{dgd} and \gls{fl} under smoothness, strong/non-convexity, and a few other technical assumptions related to the objective functions. Examples shown in \cite[Appendix A]{extended} indicate that the value of $K_{\min}$ decreases as the number of devices participating in the global update (i.e., $n$) increases.
}

\subsection{Problem Formulation} \label{sec:problemFor}
Having defined the system model and performance metrics, the next step is to design a device selection scheme that optimizes the distributed training process by minimizing its convergence time while still meeting \gls{urllc} availability requirements. 

Assuming an $n$-sync scheme and considering $d_k^{\mathrm{AI}}{\left(\bm{\Pi}_k^{\mathrm{u}}\right)}$, $K_{\min}$, and the availability requirement in equations \eqref{eq:aiTrainingDelay}, \eqref{eq:Kmin}, and \eqref{eq:availReq}, respectively, we formulate the following optimization problem:
\begin{subequations}\label{eq:opt1UserSel}
\begin{alignat}{3}
&\!\min_{\bm{\Pi}_{K_{\min}}^{\mathrm{u}}} & & \sum_{k=1}^{K_{\min}} d_k^{\mathrm{AI}}{\left(\bm{\Pi}_k^{\mathrm{u}}\right)},\tag{\ref{eq:opt1UserSel}}&\label{eq:UserSelObj}\\
&\text{s.t.}                &      &  \Pr\left\{\alpha_{i}^{\Gamma} \leq \alpha_{i}^{\mathrm{req}}\right\}\leq\gamma, \enspace\forall i \in \calU, \forall \Gamma \in \{\mathrm{UL,DL}\}, &\label{eq:UserSelC1}\\
&              &      &  \bm{1}^\intercal\bm{\pi}_k^\mathrm{u}  \geq n, \enspace\forall k \in [K_{\min}],&\label{eq:UserSelC2}\\
&              &      & \bm{\pi}_k^\mathrm{u} \in \{0,1\}^{N}, \enspace\forall k \in [K_{\min}].&\label{eq:UserSelC3}
\end{alignat}
\end{subequations}
where $\bm{1}^{\intercal}\bm{\pi}_k^\mathrm{u}{=}m_k$ and $m_k{\in}[n, N], \forall k{\in}[K_{\min}]$. 
The objective of \eqref{eq:UserSelObj} is to minimize the convergence time, which is defined as the sum of the training delays until the distributed training process achieves $\epsilon$-optimality. Constraint \eqref{eq:UserSelC1} ensures \gls{urllc} service availability, as specified by \eqref{eq:availReq}.
To fulfill the required number of local updates in the global update of \eqref{eq:dAI} in the $n$-sync scheme, \eqref{eq:UserSelC2} enforces the central node to select at least $n$ \gls{ai} devices and yet it is flexible in selecting any number of extra devices to tackle the straggling problem.
Moreover, condition \eqref{eq:UserSelC3} indicates that the device selection policy is a binary vector. Note that \eqref{eq:UserSelC2}, and \eqref{eq:UserSelC3} must be respected at all decision time epochs. 

On the one hand, the communication service availability of each \gls{urllc} device is a function of the channel state variable, $X_{i}^{\Gamma}{(t)}$, and the end-to-end delay of its packets. These two depend on many variables, such as instantaneous \gls{sinr}, selected \gls{mcs}, and transmission buffer status. On the other hand, these variables also impact the delay of the selected \gls{ai} devices, influencing the training delay for each iteration. In addition, since the \gls{urllc} service has higher priority than the distributed learning service, the amount of \gls{urllc} traffic being served on the corresponding \gls{gnb} severely affects the training delays.

The optimization problem \eqref{eq:opt1UserSel} focuses exclusively on tuning the hyper-parameters for \gls{urllc}. However, the real-time operational aspects of \gls{urllc} (e.g., transmit power decisions, retransmissions, and \gls{prb} allocation) can be further refined through a nested control loop approach. In this framework, \eqref{eq:opt1UserSel} serves as the outer control loop, while the real-time \gls{urllc} optimizations are handled by an inner control loop operating on a shorter timescale.
\section{Device Selection Optimization}\label{sec:solution}
In optimization problem \eqref{eq:opt1UserSel}, characterizing the impact of device selection policy on our \glspl{kpi} necessitates explicit modeling of the channel, queues and delay components of Equation~\eqref{eq:aiTrainingDelay}, which involves approximations that may not be accurate in practical scenarios. Therefore, we propose to model the device selection problem as a finite horizon \gls{mdp}.

In Section~\ref{sec:mdp}, we specify the state space $\calS$, action space $\calA$, and all possible rewards, $\calR$, based on dynamic interactions between the central node and \gls{ran} environment. Nevertheless, due to the complexity and dynamicity of our environment, deriving the transition probability function ($p: \calS{\times}\calR{\times}\calS{\times}\calA \mapsto [0,1]$) is impractical. Therefore, in Section~\ref{sec:sac}, we apply a model-free \gls{drl} algorithm to solve this \gls{mdp} problem. Our solution is based on \gls{bsac}~\cite{ganjTIIOrch}, an extension of the celebrated \gls{sac} algorithm~\cite{sac}. \gls{bsac} addresses the large action space challenge we observe in the optimal coexistence of \gls{urllc} and distributed learning services.

\subsection{Transformation to MDP Problem}\label{sec:mdp}
The essential elements of the \gls{mdp} are defined as follows.
\subsubsection{State Space, $\calS$} \label{sec:state}
The state space characterizes the environment. We categorize the environment's state at iteration $k$ (i.e., $\bm{s}_k\in\calS$), into three classes i) the observations from each \gls{urllc} device, ii) the observations from each \gls{ai} device, and the observations from each \gls{gnb}. In the following, we describe these three classes.

\noindent\textbf{\gls{urllc} \gls{qos} variables:}
Communication service availability of each \gls{urllc} device, as the main \gls{urllc} \gls{kpi}, is a function of \gls{per}, mean outage duration, and survival time~\cite{ganjPimrcTranslation}. Except for survival time which is static and use case specific, the state should include both (\gls{ul}/\gls{dl}) \gls{per} and (\gls{ul}/\gls{dl}) mean outage duration, estimated via empirical measurements within $\Delta t_{k-1}:= t_{k}{-}t_{k-1}$.\footnote{Notice that our two services operate on different time scales: \gls{urllc} performance should be evaluated based on the actual time, whereas the control loop's time step is defined as one iteration of the distributed learning algorithm (i.e., $d_k^{\mathrm{AI}}{\left(\cdot\right)}$ for the $k$th iteration). Hence, the control loop is not periodic in actual time and is triggered by the central node. Accordingly, we use $\Delta t_k=t_{k+1}{-}t_{k}$ to emphasize the time instants when iteration $k$ begins and ends.} In addition to these measures that explicitly affect the communication service availability, cell association, (\gls{ul}/\gls{dl}) buffer size (at $t_k$), (\gls{ul}/\gls{dl}) \gls{sinr}, and (\gls{ul}/\gls{dl}) delay are other metrics that implicitly impact the availability. However, \gls{sinr} and delay statistics might vary significantly during, possibly long, $\Delta t_{k-1}$. Hence, we represent their distribution using specific statistics of these measures, i.e., $1$st percentile, $5$th percentile and median of the \gls{sinr} distribution, and $95$th percentile, $99$th percentile, and median of the delay distribution. In fact, utilizing such percentiles is well motivated by \gls{urllc} extreme availability performance, which, under proper system design, is affected by the tail of delay and \gls{sinr} distributions~\cite{URLLCTRS}.  
     
\noindent\textbf{Distributed training delay variables:}
The training delay of each iteration, $d_k^{\mathrm{AI}}{(\bm{\Pi}_k^{\mathrm{u}})}$, is a function of the device selection; thus, we include a binary variable in the state indicating if this device has been selected in the last iteration. As indicated by \eqref{eq:aiTrainingDelay}, the delay of the \gls{dl} transmission of the global model, $d_{i,k}^{\mathrm{D}}$, and the delay of the \gls{ul} transmission of local parameters, $d_{i,k}^{\mathrm{U}}$, directly impact the training delay, and thus, should be included in the state. Besides, (\gls{ul}/\gls{dl}) buffer size (at $t_k$), and (\gls{ul}/\gls{dl}) \gls{sinr} of the underlying transmissions have an implicit effect on the delay, and therefore, we include them in the state as well. Focusing on the overall statistics (unlike \gls{urllc} service), we represent the \gls{sinr} distribution of the underlying transmissions for each \gls{ai} device with its $5$th percentile, median and $95$th percentile.
Note that no empirical measurement exists for $d_{i,k}^{\mathrm{D}}$, $d_{i,k}^{\mathrm{U}}$, and \gls{sinr} of \gls{ai} devices whose central node does not receive their local models in the $k$th iteration (i.e., $i\notin \calN_{n,k}$). Moreover, if an \gls{ai} device was selected at $t_{k-1}$, but its local update is not among the first $n$ received, its buffer size at $t_k$ would not be empty.

\noindent\textbf{\gls{gnb} level observations:}
At the \gls{gnb} level, the consumption of \glspl{prb} by each service (both in \gls{ul} and \gls{dl} directions) has a significant impact on training delay and availability. To address this, we propose adding the mean number of allocated \glspl{prb} per slot within $\Delta t_{k-1}$ for each service to the state.

\subsubsection{Action Space, $\calA$} \label{sec:actionMDP}
The action space is the set of all possible decision variables enabling the \gls{drl} agent to interact with its environment. Considering \eqref{eq:opt1UserSel}, our action vector at the $k$th iteration should be the device selection vector $\bm{\pi}_k^{\mathrm{u}}$, and thus $|\calA|=2^N$.

\subsubsection{The Reward Function, $r$} \label{sec:reward}
In general, the \gls{drl} agent follows an explicit goal, i.e., to maximize its cumulative discounted rewards. In other words, the reward function, $r_{k+1}$, is the payoff for taking action $\bm{a}_k$ at state $\bm{s}_k$. 
As \eqref{eq:availInf} indicates, the communication service availability of \gls{urllc} devices is measured in infinite time, whereas the temporal granularity of the \gls{drl} is determined by the amount of time each distributed training round takes, $\Delta t_k$. Therefore, we suggest using the availability estimator from Equation \eqref{eq:availEst} as part of the reward function. While estimating such long-term measures over a short step period might introduce inaccuracies, this estimation does represent the immediate influence of the device selection policy on the application layer performance~\cite{ganjTIIOrch}. 
A conventional approach to address constraints of \eqref{eq:opt1UserSel} uses a regularized reward function, where a penalty term is incorporated to penalize constraint violations~\cite{tesslerreward,donti2021dcreward}. Accordingly the reward for iteration $k$, $r_{k+1}$, is defined as
\begin{multline}
    r_{k+1} \coloneqq \upsilon\exp{\left(\zeta\min\left\{\min_{\substack{i\in\calU,\\\Gamma \!{\in}\! \{\mathrm{UL},\mathrm{DL}\}}}\!\!\!\!\!{\left(\hat{\alpha}_i^{\Gamma}{\left(\Delta t_{k}\right)}-\alpha_i^\mathrm{req}\right)},0\right\}\right)} \\ 
    + \left(1-\upsilon\right) \frac{T^{\mathrm{max}} - d_k^{\mathrm{AI}}{\left(\bm{\Pi}_k^{\mathrm{u}}\right)}}{T^{\mathrm{max}}},\label{eq:reward}
\end{multline}
where $\upsilon {(\in [0,1])}$ is the weight characterizing the relative importance between the \gls{urllc} and distributed learning rewards. In the \gls{urllc} reward, $\left(\hat{\alpha}_i^{\Gamma}{\left(\Delta t_k\right)}-\alpha_i^\mathrm{req}\right)$ is negative for those \gls{urllc} devices that did not meet their corresponding availability requirement within $\Delta t_k$, regardless of the transmission direction, $\Gamma$. Hence, our reward function enforces the \gls{drl} agent to maximize the availability of the worst \gls{urllc} device among those that do not meet their availability requirements. In addition, $\zeta$ is a design parameter, which is a function of the sensitivity ($\gamma$), and the precision that the maximum availability requirement needs (i.e., $\zeta$ is proportional to $\max_i\!{\left(a_i^{\mathrm{req}}\right)}{/}\gamma$).
Nevertheless, the device selection policy gets the full reward on the \gls{urllc} part (i.e., $\upsilon$) when all of the devices fulfill their availability requirements. For the distributed learning reward, the shorter $d_k^{\mathrm{AI}}\left(\cdot\right)$ is, the larger the $\bm{\pi}_k^{\mathrm{u}}$'s reward becomes. Moreover, to minimize the tail of the per-device availability distribution and the average training delay, \gls{urllc} reward decreases exponentially while the reduction in distributed learning reward is linear.

\subsection{Solution With BSAC-CoEx Framework}\label{sec:sac}
The \gls{bsac}-CoEx framework extends the \gls{sac} algorithm~\cite{sac}, integrating two additional features.
First, drawing on insights from our previous work~\cite{ganjTIIOrch}, we model the participation decision for each device using separate neurons in the last layer of the actor's \gls{dnn}, resulting in considerable scalability gains. Second, we employ a simple action masking technique to ensure that \eqref{eq:UserSelC2} is fulfilled in all decision time epochs.
Additionally, our framework enjoys the advantages of the original \gls{sac} algorithm: 
i) \gls{sac} is an off-policy model-free \gls{drl} algorithm in which explorations aim to identify an optimal policy that maximizes not only the expected accumulated discounted reward but also the expected entropy at each visited state, thus improving the stability and exploration~\cite{sac},
ii) \gls{sac} has an actor-critic architecture where the critic approximates state-action pair values, and the actor determines the policy, and 
iii) SAC conquers the massive sampling complications and minimizes the sensitivity to its hyperparameters. In the rest of this section, we first introduce \gls{sac} for our problem, followed by a detailed description of \gls{bsac}-CoEx as implemented for our device selection problem.

The main objective in \gls{sac} is to find an optimal stochastic policy that maximizes the discounted sum of reward and entropy. 
Thus, the objective of \gls{sac} is
\begin{align}\label{eq:sacEntropy}
\hspace{-1.5mm}\pi^\star{\left(\cdot|\cdot\right)} {\coloneq} \argmax_{\pi\left(\cdot|\cdot\right)}\!\! \sum_{k=1}^{K_{\min}}\mathbb{E}{\left[\sum_{\kappa=k}^{K_{\min}{-}1}\!\!\!\lambda^{\kappa-k} \mathbb{E}{\left[r_{\kappa{+}1}{+}\psi \mathbb{H}{\left(\pi\left(\cdot|\bm{s}_\kappa\right)\right)}\right]}\right]},   
\end{align}
where $\lambda {(\in[0,1])}$ and $\psi{(>0)}$  are the discount factor and the temperature parameter specifying the relative importance between the reward and entropy terms, respectively~\cite{sac}. Furthermore, $\mathbb{H}{\left(\pi\left(\cdot|\bm{s}_k\right)\right)} \coloneqq \mathbb{E}{\left[-\ln\!{\left(\pi{(\bm{a}|\bm{s}_k)}\right)}\right]}$ is the entropy of policy $\pi$ at state $\bm{s}_k$. Introducing entropy in \eqref{eq:sacEntropy} guides the policy to explore more broadly while avoiding blatantly unfavorable trajectories.


The \gls{sac} algorithm (and actor-critic methods in general) uses policy iteration, in which the algorithm alternates between policy evaluation (to compute the state-action value function by $Q_{\pi}{(\bm{s},\bm{a})}$) and policy improvement (to compute $\pi$) in the direction of maximizing the sum of discounted return (i.e., sum of reward and a portion of entropy here). 
In the policy evaluation step, using the soft Bellman backup equations, the soft state-action value function can be computed iteratively as follows~\cite{sac}:
\begin{align}
Q_{\pi}{(\bm{s}_{k},\bm{a}_{k})} = r_{k+1} {+} \lambda\mathbb{E}{\left[Q{\left(\bm{s}_{k+1},\bm{a}_{k+1}\right)}{+}\psi \mathbb{H}{\left(\pi\left(\cdot|\bm{s}_k\right)\right)}\right]}. \label{eq:sacQ}
\end{align}

In large-scale reinforcement learning problems where the state and action spaces are large, $Q_{\pi}$ and $\pi$ are approximated in each iteration using \glspl{dnn} (via critics and actors, respectively). In \gls{bsac}-CoEx, we leverage target networks and clipped double Q-learning. Hence, in our architecture, there are $4$ \glspl{dnn} for the first critic, its target critic, the second critic, and its target, whose weights are denoted by $\bm{\varphi}_1$, $\bm{\tilde{\varphi}}_1$, $\bm{\varphi}_2$, and $\bm{\tilde{\varphi}}_2$, respectively. In addition, there are $2$ \glspl{dnn} for the actor and its target network, whose weights are denoted by $\bm{\vartheta}$ and $\bm{\tilde{\vartheta}}$, respectively. Thus, $Q_{\pi}$, in \eqref{eq:sacQ}, is approximated by $2$ \glspl{dnn} as $Q_{\bm{\varphi}_i}$, $i{\in}\{1,2\}$. 
While \gls{sac} can handle both discrete and continuous actions, accommodating our action space with discrete actions requires the last layer of the actor networks to have as many neurons as the size of the action space, which leads to impractical memory and computational time requirements~\cite{ganjTIIOrch}. 
An alternative is to employ a single continuous action that is subsequently quantized to a discrete value, denoting the device participation for all \gls{ai} devices.
Nevertheless, due to a large action space and, thus, an enormous set of quantized values, the performance remains unsatisfactory.
To overcome these limitations, in \gls{bsac}-CoEx, we represent $\bm{a}_k$ as a continuous vector. As discussed in Section~\ref{sec:actionMDP}, our action space $\calA$ has $N$ dimensions for our device selection problem. 
Thus, each element $[\bm{a}_k]_i{\in}\calA_i, \forall i{\in}\calN$ indicates the action for the $i$th \gls{ai} device, with $\calA_i$ denoting the action space of the $i$th dimension, which corresponds to the action space of the $i$th \gls{ai} device. 
To eliminate discrete binary actions, we represent $\calA_i$ as continuous, and for sampling reasons described later, all elements are confined within $\tanh$ bounds (i.e., $[\bm{a}_k]_i{\in}\left[-1,1\right], \forall i{\in}\calN$). 
With this approach, the size of the actor networks' final layer increases linearly with the number of \gls{ai} devices, avoiding the exponential growth associated with a discrete action space. Moreover, to mask out selections that do not follow condition \eqref{eq:UserSelC2}, we determine the mapping from $\bm{a}_k$ to $\bm{\pi}_k^{\mathrm{u}}$ by
\begin{equation} \label{eq:actionToselection}
  \bm{\pi}_k^{\mathrm{u}}\coloneq\begin{cases}
    \mathds{1}\{\bm{a}_k\succeq \bm{0}\}, & \text{if $\bm{1}^\intercal\mathds{1}\{\bm{a}_k\succeq \bm{0}\}\geq n$},\\
    \mathds{1}\{\bm{a}_k\succeq a_{k}^{(n)}\bm{1}\}, & \text{otherwise},
  \end{cases}
\end{equation}
where $a_{k}^{(n)}$ is the $n$th largest element in vector $\bm{a}_k$.

Let us assume the transitions are stored in a replay buffer, $\calB$. Then, regardless of sampling technique (e.g., uniform or prioritized experience replay) and the mini-batch size ($|\calB_{\mathrm{mb}}|$), a sampled transition can be represented with ${(\bm{s},\bm{a},r,\hat{\bm{s}}, I)}$, where $I$ is a binary parameter that is $0$ if the distributed learning converges in $\hat{\bm{s}}$, and is $1$, otherwise.
Then, $\bm{\varphi}_1$ and $\bm{\varphi}_2$ can be trained by minimizing the \gls{mse} for each sampled transition as
\begin{align}
J_\mathrm{Q}(\bm{\varphi}_i)\coloneq\mathbb{E}{\left[\frac{1}{2}\!\left(Q_{\bm{\varphi}_i}\!{(\bm{s},\bm{a})}{-}\widetilde{Q}{(\hat{\bm{s}},\tilde{\bm{a}},r,I)}\right)^2\right]},  \label{eq:lossQJ}
\end{align}
where $\tilde{\bm{a}}$ is sampled from $\pi_{\tilde{\vartheta}}{(\cdot|\bm{s}_{k+1})}$, and 
\begin{align}
\widetilde{Q}{(\bm{s},\bm{a},r,I)} \coloneq r + I \lambda {\left(\min_{i=1,2} Q_{\bm{\tilde{\varphi}}_i}{\left(\bm{s},\bm{a}\right)}-\psi \ln\!{\left(\pi{(\bm{a}|\bm{s})}\right)}\right)}.   \label{eq:sacTargetQ}
\end{align}
Note that the minimum represents the smallest Q-value between the two state-action value function approximations for clipped double Q-learning~\cite{td3}.
Then, in order to minimize $J_Q{(\cdot)}$, $\bm{\varphi}_1$ and $\bm{\varphi}_2$ are updated in the direction of gradient descent.
To ensure that temporal-difference error remains low, we update target critics' weights gradually by $\bm{\tilde{\varphi}}_i {=} \nu\bm{\varphi}_i{+}(1{-}\nu)\bm{\tilde{\varphi}}_i$ for $i{\in}\{1,2\}$ at each \gls{drl} iteration.

In the policy improvement step, the actor \gls{dnn} can be updated by minimizing the expected Kullback-Leibler divergence between $\pi_{\vartheta}$ and the exponential of the soft state-action value function, which can be rewritten as
\begin{align}\label{eq:lossPolicy1}
J_\mathrm{\pi}{(\bm{\vartheta})}\coloneq\mathbb{E} \left[\psi\ln\!{\left(\pi_{\bm{\vartheta}}{\left(\!\bm{a}|\bm{s}\right)}\right)}-\min_{i=1,2} Q_{\bm{\varphi}_i}\!{\left(\!\bm{s},\bm{a}\right)}\right].
\end{align}

To minimize $J_\mathrm{\pi}{(\cdot)}$ based on the latest policy, we employ the re-parameterization technique, from~\cite{sac}, to reformulate the expectation over actions into an expectation over noise, leading to a smaller variance estimator.
Therefore, we draw samples from an invertible $\tanh$ function of Gaussian policy such that $\widehat{\bm{a}}{\coloneq}\tanh{\left(\mu_{\mathrm{G},\bm{\vartheta}}{(\bm{s})}+\sigma_{\mathrm{G}, \bm{\vartheta}}{(\bm{s})}\cdot\bm{\chi}\right)}$, where $\mu_{\mathrm{G},\bm{\vartheta}}{(\cdot)}$ and $\sigma_{\mathrm{G}, \bm{\vartheta}}{(\cdot)}$ are the estimated mean and standard deviation of a Gaussian distribution, respectively, and $\bm{\chi}$ follows a multivariate Gaussian distribution, with a mean of $\bm{0}$, and an identity matrix as its covariance matrix. Accordingly, we can reformulate $J_{\mathrm{\pi}}{(\bm{\vartheta})}$ by replacing $\bm{a}$ with $\widehat{\bm{a}}$ in \eqref{eq:lossPolicy1}.
The policy parameters, $\bm{\vartheta}$, are then updated in the gradient descent direction as in~\cite{sac}. Additionally, we update the target actor weights smoothly by $\bm{\tilde{\vartheta}} = \nu\bm{\vartheta}+(1-\nu)\bm{\tilde{\vartheta}}$.

\begin{algorithm}[t]\footnotesize 
	\SetKwFor{ForPar}{for}{do in parallel}{end}
	\SetAlgoLined
	\textbf{Input}: Set of \gls{ai} devices $\calN$, Required number of local models in global update $n$\;
	\textbf{Output}: Device selection policy as a function of \gls{ran} state\;
	\textbf{Initialize}: $\bm{\varphi}_1$, $\bm{\varphi}_2$, and $\bm{\vartheta}$ and set $\bm{\tilde{\varphi}}_1\gets\bm{\varphi}_1$, $\bm{\tilde{\varphi}}_2\gets\bm{\varphi}_2$, and $\bm{\tilde{\vartheta}}\gets\bm{\vartheta}$, $k\gets1$\;
	\Comment*[h]{episode: numbers of iterations during which distributed learning converges}\\
	\ForEach{episode}{
	    Set initial device selection, from previous virtual training or random\;
	    \While{true}{
    		Receive $\bm{c}_{i,k}$ from $n$ \gls{ai} devices, or $T^{\max}$\;
    		Observe $\bm{s}_k$ (measured within $\Delta t_{k-1}$)\;
    		\lIf{$d_k^{\mathrm{AI}} < T^{\max}$}{Compute $\bw_{k+1}$ as in \eqref{eq:globalUp}}
    		\lElse{$\bw_{k+1}\gets \bw_{k}$}
    		Sample an action, $\bm{a}_k {\sim} \pi_{\vartheta}{(\cdot|\bm{s}_{k})}$\;
    		Transmit $\bw_{k+1}$, via \gls{ran}, to selected \gls{ai} devices ($\bm{\pi}_{k}^{\mathrm{u}}$, derived from \eqref{eq:actionToselection})\;
    		Observe $\bm{s}_{k+1}$, and calculate $r_{k+1}$ via \eqref{eq:reward}\;
    		\lIf{distributed learning converges as in \eqref{eq:Kmin}}{$I_{k+1}\gets0$}\lElse{$I_{k+1}\gets1$}
    		Store ${\left(\bm{s}_k, \bm{a}_k, r_{k+1}, \bm{s}_{k+1}, I_{k+1}\right)}$ in $\calB$\;
    		\lIf{$I_{k+1}=0$}{break}
    		$k\gets k+1$\;
	    }
	   \Comment*[h]{Training \glspl{dnn}}\\
	    \If{$|\calB|{\geq}|\calB|_{\min}$ \& $k = l|\calB_\mathrm{mb}|, \forall l{\in}\{1,2,\ldots\}$}{
            Randomly sample a mini-batch $\calB_{\mathrm{mb}}$ from $\calB$\;
            \ForAll{${\left(\bm{s}, \bm{a}, r, \hat{\bm{s}}, I\right)}\in\calB_{\mathrm{mb}}$}{
                Derive $\widehat{Q}$ using \eqref{eq:sacTargetQ}, where $\tilde{\bm{a}}{\sim}\pi_{\tilde{\vartheta}}{(\cdot|\hat{\bm{s}})}$\;
            }
            $\bm{\varphi}_i \gets \bm{\varphi}_i {-} \frac{1}{|\calB_\mathrm{mb}|}\sum_{\forall b\in\calB_{\mathrm{mb}}}\!\!\widehat{\nabla}J_\mathrm{Q}(\bm{\varphi}_i)$, for $i{\in}\{1,2\}$\;
            $\bm{\vartheta}{\gets}\bm{\vartheta} {-} \frac{1}{|\calB_\mathrm{mb}|} \!\sum_{\forall b\in\calB_{\mathrm{mb}}}\!\!\!\widehat{\nabla}\!J_{\!\mathrm{\pi}}(\bm{\vartheta})$, where \eqref{eq:lossPolicy1} uses $\widehat{\bm{a}}$\;
            $\bm{\tilde{\varphi}}_i \gets \nu\bm{\varphi}_i{+}(1{-}\nu)\bm{\tilde{\varphi}}_i$ for $i{\in}\{1,2\}$\;
            $\bm{\tilde{\vartheta}} \gets \nu\bm{\vartheta}+(1-\nu)\bm{\tilde{\vartheta}}$\;
        }
	}
\caption{BSAC-CoEx Algorithm}
\label{alg:alg1}
\end{algorithm}
\algorithmcfname\,\ref{alg:alg1} summarizes the learning procedure of \gls{bsac}-CoEx. Leveraging the off-policy learning capability of such a framework, one can train the \glspl{dnn} via either the virtual network (e.g., digital twin or realistic simulations) or an operating network (e.g., in safe exploration mode). On the former, $\bm{a}_k$ can be sampled using the behavior policy in each episode, and episodes can run in parallel to speed up the training. Nevertheless, on the operating network, this algorithm can switch to on-policy learning (i.e., $\bm{a}_k$ is sampled via the most updated policy, and episodes run consecutively). A hybrid strategy in which the \glspl{dnn} are trained first with a virtual network and then tuned via an operational network could potentially result in a more efficient learning procedure. 

\algorithmcfname\,\ref{alg:alg1} was initially developed for single-task learning within a star topology, where computation occurs on a single server. However, minor modifications can alleviate both assumptions. 
First, in a multi-task learning scenario, the communication load among \gls{ai} devices may increase without affecting the overall communication architecture and the skeleton of \algorithmcfname\,\ref{alg:alg1}. 
Nevertheless, it is necessary to adjust the definitions of parameters involved in the optimization problem \eqref{eq:opt1UserSel}. For instance, when training one independent model per task, $K_{\min}$ in \eqref{eq:opt1UserSel} should be set to the maximum $K_{\min}$ value across all individual tasks. Additionally, $d_k^{\mathrm{AI}}{(\cdot)}$ should take into account the increased delay caused by communicating multiple models per iteration.
Second, for other computation topologies (e.g., from a single parameter server to a multi-parameter server scenario to multi-tier tree topology), we need to account for extra communication overhead for parameter exchange among parameter servers. Again, notation-wise, \eqref{eq:opt1UserSel} and \algorithmcfname\,\ref{alg:alg1} remain unchanged, but we need to adjust the definitions of parameters characterizing $K_{\min}$ and $d_k^{\mathrm{AI}}{(\cdot)}$.

In the following section, we describe our simulator's modeling principles and its configuration.
\section{Simulation Methodology and Configuration}\label{sec:simulation}
For simulating the deployment where \gls{urllc} and distributed learning services coexist, we considered a factory automation scenario, which is one of the main use cases of \gls{urllc} services~\cite{urllcApp} (demonstrated in \figurename\,\ref{fig:simulationSetup}). More explicitly, we designed a 3D model of a small factory of size $40\times40\times10$\,m$^3$ with $4$ \glspl{gnb} at the height of $8$\,m, and with an inter-site distance of $20$\,m.
\begin{figure}[t]
	\centering
	\includegraphics[width=.9\columnwidth,keepaspectratio]{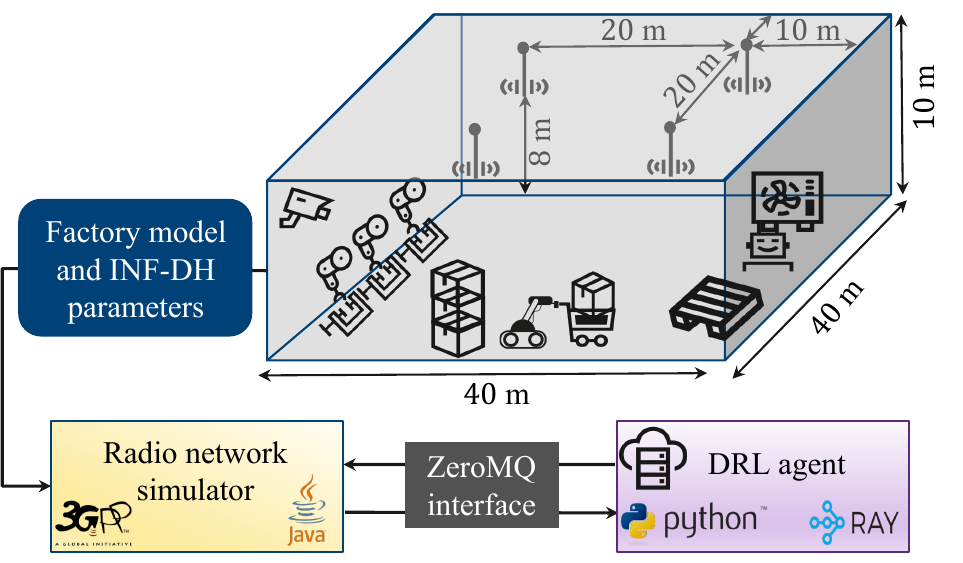}
	\caption{The simulation setup.}
	\label{fig:simulationSetup}
	\vspace{-6mm}
\end{figure}

\subsection{InF-DH Scenario}\label{sec:simulationCm}
In our simulations, we employed the base \gls{3gpp} channel model 
\ifthenelse{\boolean{appendixversion}}{
of Appendix~\ref{app:channelModel}
}{
\cite[Appendix B]{extended}
}for the \gls{inf-dh} scenario, where the \gls{gnb} and the devices are placed, respectively, higher and under the average height of the clutters~\cite{3GPP38901}. Details on these parameters can be found in
\ifthenelse{\boolean{appendixversion}}{
Appendix~\ref{app:inf}.
}{
\cite[Appendix C]{extended}.
}

\subsection{Radio Network Simulator and DRL Agent}

The radio network simulator is event-based, \gls{3gpp} compliant, and operates at \gls{ofdm} symbol resolution. We considered numerology one from~\cite{dahlman5GNr}, implying that each slot and symbol are $0.5$\,ms and $33.33$\,$\mu$s long, respectively. 
\ifthenelse{\boolean{appendixversion}}{
We assumed the channel response matrix in \eqref{eq:channel} remains constant during one slot.
} 
We configured \gls{rlc} in \gls{am} for distributed learning, ensuring a seamless training process throughout the simulation. In contrast, \gls{rlc}'s slow retransmissions are of minimal benefit to \gls{urllc} packets, which demand low latency~\cite{dahlman5GNr}. Thus, we set \gls{rlc} in \gls{um} for \gls{urllc} flow.
Within \gls{urllc} flow, \gls{ul} and \gls{dl} \gls{urllc} traffic are scheduled based on round robin and delay, respectively. Here, the delay-based scheduling prioritizes the packet that has been in the queue the longest. Moreover, we used proportional-fair scheduling for distributed training traffic in both directions. 
However, we assumed the \gls{urllc} flow has strict priority over the distributed learning flow, implying that packets generated by the distributed learning service would only be scheduled when no \gls{urllc} packets are in the queues.

During transmission, the \gls{mac} layer picks a number of packets from the respective \gls{rlc} buffer, depending on the selected \gls{mcs} on lower layers. Alternatively, \gls{rlc} can segment packets into smaller pieces to accommodate them into transport blocks for transmission. At the receiver, the instantaneous \gls{sinr} of each transport block, influenced by the radio channel and dynamic interference of other devices' transmissions, dictates an error probability. Following this, the receiving \gls{rlc} entity reassembles successfully decoded segments and forwards them to the application layer. When assessing availability at the application layer, a \gls{urllc} packet is deemed lost if not entirely received within its delay bound. The application state is then determined, integrating $T_{\mathrm{sv}}$ according to \eqref{eq:appState}. Finally, communication service availability is computed using \eqref{eq:availEst}. 

The \gls{urllc} traffic is represented by periodic \gls{ul} and \gls{dl} traffic, with delay thresholds of $6$\,ms and $4$\,ms and payload sizes of $64$ bytes and $80$ bytes, respectively, both with a period of $6$\,ms. Such \gls{urllc} traffic characterization aligns with the machine control use case for mobile robots in~\cite{3GPP22104}.
Motivated by~\cite{3GPP22874AiModel}, we assumed that the shared \gls{dnn} architecture, employed both on devices and the central node, follows MobileNets~\cite{mobileNets}. MobileNets are a class of efficient \gls{dnn} models designed for mobile and embedded vision applications based on a simplified architecture. We considered $0.25$\,MobileNet-$224$ in~\cite{mobileNets}, implying that the \gls{dnn} model has $0.5$ million parameters.
To model the distributed training traffic, we assumed \gls{fl} and 32 bits quantization for each model parameter (i.e., each model, whether local or global, can be represented as a $2$\,MB packet). Nevertheless, our solution applies to settings where other quantization/compression approaches reduce the communication overhead~\cite{Sindri2020Linear}.

In our simulation setup, the \gls{drl} agent is located in a separate server and interacts with the radio network simulator through a ZeroMQ interface. We ran the \gls{drl} agent on a server with Intel(R) Xeon(R) Gold 6132 CPU @ 2.60\,GHz, $8$ cores and $64$\,GB of RAM. In the exploration phase, we trained the \gls{drl} agent for $7\,000$ episodes of $50$-iteration length, and our simulation time differed depending on how long each iteration (i.e., $d_k^{\mathrm{AI}}(\cdot)$) took. Nevertheless, our \gls{bsac}-CoEx converged with significantly fewer iterations at around $150\,000$ iterations. \tablename\,\ref{tab:simSetup} outlines the simulation parameters.

\begin{table}[t]
\footnotesize
\centering
\caption{Simulation Parameters}
\label{tab:simSetup}
\renewcommand{\arraystretch}{1.1}
\begin{tabular}{l||l}
	\hlineB
	\multicolumn{2}{c}{\textbf{Radio Network Simulator Parameters}} \\
	\hline
	\textbf{Parameter}& \textbf{Value}\\
	\hline
	Deployment   & $4$ \glspl{gnb}\\
	Duplex/Carrier frequency   & FDD/$2.6$\,GHz\\
	\gls{gnb} antenna height&   $8$\,m \\
	Devices' height &$1.5$\,m \\
	Number of antenna elements in \gls{gnb}/device & $2$/$2$\\ 
	Bandwidth&   $40$\,MHz\,\\
	TTI length/Subcarrier spacing& $0.5$\,ms/$30$\,KHz  \\
	UL/DL transmit power &   $0.2$\,W/$0.5$\,W  \\
	Max\,num\,of\,\gls{ul}/\gls{dl}\,\gls{urllc}\,Trans. (\gls{mac}) & $3/2$ \\
	Max\,num\,of\,\gls{ul}/\gls{dl}\,\gls{ai}\,Trans. (\gls{mac}) & $10/10$\\
	Max\,num\,of\,\gls{ul}/\gls{dl}\,\gls{ai}\,Trans. (\gls{rlc}) & $8/8$ \\
	\gls{ul}/\gls{dl}\,\gls{urllc}\,delay bound& $6/4$\,ms \\
	\gls{ul}/\gls{dl} \gls{urllc} survival time, $T_{\mathrm{sv}}$& $6/6$\,ms \\
	Total number of \gls{ai} devices, $N$& $50$ \\
	Req\,num\,of\,local models to proceed in \eqref{eq:globalUp}, $n$ & $15$\\
	$\upsilon$/$\zeta$, in the reward function \eqref{eq:reward} & $0.5$/$100$\\
        $T^{\max}$, in \eqref{eq:aiTrainingDelay}& $10$\,s \\
	\hline
	\multicolumn{2}{c}{\textbf{DRL Agent Parameters}} \\
	\hline
	\textbf{Parameter}& \textbf{Value}\\
	\hline
    Replay buffer size & $1\,000\,000$\\
	Training mini-batch size, $|\calB_\mathrm{mb}|$ & $200$\\
    Discount factor, $\lambda$   & $0.1$\\
	Neural network hidden layers (all six) & $128\times128$\\
	Prioritized replay buffer $\alpha$/$\beta$ in~\cite{prioReplay} & $0.6$/$0.4$\\
	Learning rate (for critic, actor and entropy)   & $0.0003$ \\
	$\nu$ (for smooth update) & $0.002$\\
	\hlineB
\end{tabular}
\end{table}

In the following section, we run comprehensive simulations to analyze the impact of various design parameters, including distributed training load, \gls{urllc} traffic, the number of selected \gls{ai} devices, and how network resources are sliced between \gls{urllc} and distributed learning services.
\section{Results and Discussion}\label{sec:preformance}

For the performance evaluation, we set up two benchmarks.
\begin{enumerate}
    \item Semi-random \gls{urllc} devices: In this benchmark, we set up $10$ \gls{urllc} devices that move in a 1D space at a speed of $30$\,km/h, while remaining in close proximity to a fixed position that is maintained in different simulations. Yet, the movement direction of each device is randomly changed in different seeds.
    \item Random \gls{urllc} devices: In this benchmark, we set up $20$ \gls{urllc} devices. At each simulation, the \gls{urllc} devices appear at random positions and move in 1D in a random direction at a speed of $30$\,km/h within a short distance of that position.
\end{enumerate}
The results of the first benchmark highlight how channel conditions and interference affect device selection, while those of the second benchmark provide insights into long-term system-level performance.
For each of these benchmarks, we compared our solution (shown as \ours{} in figures) against the following baselines: 
\begin{itemize}
    \item \su{}: We did not have any \gls{ai} device in the factory, and the \gls{urllc} devices could consume the entire $40$\,MHz bandwidth and $0.5$\,W \gls{dl} transmit power.

    \item \mixr{m}: It involved both \gls{urllc} and \gls{ai} devices sharing the wireless resources. We randomly selected a set of $m$ participating devices (i.e., $m_k{=}m, \forall k{\in}\mathbb{N}$), where $m{\in}\{15,20,30,40,50\}$.
        
    \item \slr{m}: We dedicated $25\%$ of resources to the \gls{urllc} service (i.e., $10$\,MHz bandwidth and $0.125$\,W for total \gls{dl} transmit power), and the rest to the distributed learning service.\footnote{In our experiment setup, we observed that allocating less than $25\%$ of resources to \gls{urllc} traffic substantially increases the violation probability (formulated in \eqref{eq:availReq}), making them unsuitable for the factory automation use cases.} Besides, we randomly picked a set of $m$ participating devices (i.e., $m_k{=}m, \forall k{\in}\mathbb{N}$), where $m{\in}\{15,20\}$.

    \item \mixo{}: This baseline is inspired by references \cite{energyDelayEarlyAccessTmcTd3,deviceSelMab,energyDelayTiiDdpg} which minimize the convergence delay of federated learning via device selection, unaware of the potential coexisting services.
    As in \mixr{m}, the wireless resources are shared in this baseline too. Device selection was executed using \gls{bsac},\footnote{We had to replace the \gls{drl} solution algorithm with \gls{bsac} for memory and computation purposes. Convergence and scalability results of \gls{bsac} are out of the scope of this paper. Interested readers are referred to~\cite{ganjTIIOrch}.} ignoring the \gls{urllc} service representation. Thus, in this baseline, the state only contained distributed training delay variables and its \gls{prb} usage, and the reward in \eqref{eq:reward} was confined to the distributed learning part. 
    
    \item \slo{}: We sliced the bandwidth and transmit power as in \slr{m}. However, the device selection was performed as in \mixo.
    
\end{itemize}
The \su{} baseline represents the best possible performance on the \gls{urllc} availability in our scenario.
For those baselines involving distributed learning service, we set $N{=}50$ and $n{=}15$
In \ours{}, and to calculate the reward in \eqref{eq:reward}, we set $\upsilon$ and $\zeta$ to $0.5$, and $100$, respectively.
Besides, we assumed that all \gls{urllc} devices serve a single use case, and thus, set the availability requirement to $0.99$ (i.e., $\alpha_i^\mathrm{req}{=}\alpha^\mathrm{req}{=}0.99, \forall i{\in}\calU$). 
For \su{} evaluations, we conducted $102$-second simulations $300$ times with different seeds. The rest of the baselines were evaluated with at least $300$ simulations of $50$ iterations each, resulting in different simulation lengths. 
Note that there is no progress in distributed training if all $n$ local models are not collected by the central node within a duration of $T^{\max}$, and thus, \gls{drl} iterations may vary from distributed training iterations. Regardless, if there is no strict latency constraint from the distributed learning task, we can tune $T^{\max}$ sufficiently large to ensure that time-outs rarely happen. In our evaluations, we set $T^{\max}{=}10$\,s. 

\begin{figure*}[!t]
\centering
    \begin{subfigure}[b]{0.49\textwidth}
        \centering {\footnotesize\input{./Components/Figs/journal/availCdfV2}}
        \caption{}
        \label{fig:availUrllc}
    \end{subfigure}
    \hfill
    \begin{subfigure}[b]{0.49\textwidth}
        \centering
        {\footnotesize
%
%
\definecolor{mycolorRand7}{rgb}{0.12, 0.3, 0.17}
\definecolor{medianColor}{rgb}{0.7, 0.11, 0.11}

\definecolor{mycolorSlic}{rgb}{0.45, 0.31, 0.59}
\definecolor{mycolorSlic}{rgb}{0.5, 0.0, 0.5}
\definecolor{mycolorSlic}{rgb}{0.47, 0.32, 0.66}

\definecolor{mycolorDrl}{rgb}{0.0, 0.0, 0.55}
\definecolor{mycolorDrl}{rgb}{0.0, 0.14, 0.4}
\definecolor{mycolorDrl}{rgb}{0.25, 0.41, 0.88}
\begin{tikzpicture}
\begin{axis}[%
width=72.0*0.99mm,
height=46.0mm,
at={(0.769in,0.482in)},
scale only axis,
unbounded coords=jump,
xmin=0.5,
xmax=8.5,
xtick={1,2,3,4,5,6,7,8},
xticklabels={{$m_k$},{$15$},{$20$},{$15$},{$20$},{$30$},{ $40$},{$50$}},
ymin=0.8,
ymax=10.05,
xlabel style={font=\color{white!15!black},yshift=-0.7mm},
xlabel={Number of selected AI devices},
ylabel style={font=\color{white!15!black}},
ylabel={Training delay (s)},
axis background/.style={fill=white},
xmajorgrids,
ymajorgrids,
legend style={at={(0.01,0.96)}, anchor=north west, legend cell align=left, align=left, draw=none,fill=white, inner sep=0.001pt, outer sep=0.001pt, row sep=0pt}
]
\addplot [color=mycolorDrl,line width=0.75pt, solid, forget plot]
  table[row sep=crcr]{%
1	1.35\\
1	2.584\\
};  
\addplot [color=mycolorSlic, line width=0.75pt, solid, forget plot]
  table[row sep=crcr]{%
2	2.962\\
2	10\\
};
\addplot [color=mycolorSlic, line width=0.75pt, solid, forget plot]
  table[row sep=crcr]{%
3	3.8435\\
3	5.93700000000001\\
};
\addplot [color=mycolorRand7, line width=0.75pt, solid, forget plot]
  table[row sep=crcr]{%
4	2.2315\\
4	10\\
};
\addplot [color=mycolorRand7, line width=0.75pt, solid, forget plot]
  table[row sep=crcr]{%
5	2.8685\\
5	4.526\\
};
\addplot [color=mycolorRand7, line width=0.75pt, solid, forget plot]
  table[row sep=crcr]{%
6	4.02700000000001\\
6	6.00349999999997\\
};
\addplot [color=mycolorRand7, line width=0.75pt, solid, forget plot]
  table[row sep=crcr]{%
7	3.884\\
7	10\\
};
\addplot [color=mycolorRand7, line width=0.75pt, solid, forget plot]
  table[row sep=crcr]{%
8	4.2035\\
8	6.26400000000001\\
};
\addplot [color=mycolorDrl, line width=0.75pt, solid,  forget plot]
  table[row sep=crcr]{%
1	0.914999999999999\\
1	1.1985\\
};
\addplot [color=mycolorSlic, line width=0.75pt, solid, forget plot]
  table[row sep=crcr]{%
2	1.5355\\
2	2.3975\\
};
\addplot [color=mycolorSlic, line width=0.75pt, solid, forget plot]
  table[row sep=crcr]{%
3	1.8295\\
3	3.1105\\
};
\addplot [color=mycolorRand7, fill=mixedRandColor2, line width=0.75pt, solid, forget plot]
  table[row sep=crcr]{%
4	1.126\\
4	1.79625\\
};
\addplot [color=mycolorRand7, fill=mixedRandColor2, line width=0.75pt, solid, forget plot]
  table[row sep=crcr]{%
5	1.4285\\
5	2.3005\\
};
\addplot [color=mycolorRand7, fill=mixedRandColor2, line width=0.75pt, solid, forget plot]
  table[row sep=crcr]{%
6	1.629\\
6	3.084\\
};
\addplot [color=mycolorRand7, fill=mixedRandColor2, line width=0.75pt, solid, forget plot]
  table[row sep=crcr]{%
7	1.674\\
7	3.312375\\
};
\addplot [color=mycolorRand7, fill=mixedRandColor2, line width=0.75pt, solid, forget plot]
  table[row sep=crcr]{%
8	1.811\\
8	3.7295\\
};
\addplot [color=mycolorDrl, line width=0.75pt, solid, forget plot]
  table[row sep=crcr]{%
0.875	2.584\\
1.125	2.584\\
};
\addplot [color=mycolorSlic, line width=0.75pt, solid,  forget plot]
  table[row sep=crcr]{%
1.875	10\\
2.125	10\\
};
\addplot [color=mycolorSlic, line width=0.75pt, solid,  forget plot]
  table[row sep=crcr]{%
2.875	5.93700000000001\\
3.125	5.93700000000001\\
};
\addplot [color=mycolorRand7, fill=mixedRandColor2, line width=0.75pt, solid, forget plot]
  table[row sep=crcr]{%
3.875	10\\
4.125	10\\
};
\addplot [color=mycolorRand7, fill=mixedRandColor2, line width=0.75pt, solid, forget plot]
  table[row sep=crcr]{%
4.875	4.526\\
5.125	4.526\\
};
\addplot [color=mycolorRand7, fill=mixedRandColor2, line width=0.75pt, solid, forget plot]
  table[row sep=crcr]{%
5.875	6.00349999999997\\
6.125	6.00349999999997\\
};
\addplot [color=mycolorRand7, fill=mixedRandColor2, line width=0.75pt, solid, forget plot]
  table[row sep=crcr]{%
6.875	10\\
7.125	10\\
};
\addplot [color=mycolorRand7, fill=mixedRandColor2, line width=0.75pt, solid, forget plot]
  table[row sep=crcr]{%
7.875	6.26400000000001\\
8.125	6.26400000000001\\
};
\addplot [color=mycolorDrl, line width=0.75pt, solid, forget plot]
  table[row sep=crcr]{%
0.875	0.914999999999999\\
1.125	0.914999999999999\\
};
\addplot [color=mycolorSlic, line width=0.75pt, solid,  forget plot]
  table[row sep=crcr]{%
1.875	1.5355\\
2.125	1.5355\\
};
\addplot [color=mycolorSlic, line width=0.75pt, solid,  forget plot]
  table[row sep=crcr]{%
2.875	1.816\\
3.125	1.816\\
};
\addplot [color=mycolorRand7, fill=mixedRandColor2, line width=0.75pt, solid, forget plot]
  table[row sep=crcr]{%
3.875	1.126\\
4.125	1.126\\
};
\addplot [color=mycolorRand7, fill=mixedRandColor2, line width=0.75pt, solid, forget plot]
  table[row sep=crcr]{%
4.875	1.4285\\
5.125	1.4285\\
};
\addplot [color=mycolorRand7, fill=mixedRandColor2, line width=0.75pt, solid, forget plot]
  table[row sep=crcr]{%
5.875	1.629\\
6.125	1.629\\
};
\addplot [color=mycolorRand7, fill=mixedRandColor2, line width=0.75pt, solid, forget plot]
  table[row sep=crcr]{%
6.875	1.674\\
7.125	1.674\\
};
\addplot [color=mycolorRand7, fill=mixedRandColor2, line width=0.75pt, solid, forget plot]
  table[row sep=crcr]{%
7.875	1.811\\
8.125	1.811\\
};
\addplot [color=mycolorDrl, fill=ourSolColor, line width=.25pt, solid, area legend]
  table[row sep=crcr]{%
0.75	1.1985\\
0.75	1.35\\
1.25	1.35\\
1.25	1.1985\\
0.75	1.1985\\
};
\addplot [color=mycolorSlic, fill=slicingColor, line width=0.4pt, area legend]
  table[row sep=crcr]{%
1.75	2.3975\\
1.75	2.962\\
2.25	2.962\\
2.25	2.3975\\
1.75	2.3975\\
};
\addplot [color=mycolorSlic, fill=slicingColor, line width=0.4pt, forget plot]
  table[row sep=crcr]{%
2.75	3.1105\\
2.75	3.8435\\
3.25	3.8435\\
3.25	3.1105\\
2.75	3.1105\\
};
\addplot [color=mycolorRand7, fill=mixedRandColor2, line width=0.4pt, solid, area legend]
  table[row sep=crcr]{%
3.75	1.79625\\
3.75	2.2315\\
4.25	2.2315\\
4.25	1.79625\\
3.75	1.79625\\
};
\addplot [color=mycolorRand7, fill=mixedRandColor2, line width=0.4pt, solid, forget plot]
  table[row sep=crcr]{%
4.75	2.3005\\
4.75	2.8685\\
5.25	2.8685\\
5.25	2.3005\\
4.75	2.3005\\
};
\addplot [color=mycolorRand7, fill=mixedRandColor2, line width=0.4pt, solid, forget plot]
  table[row sep=crcr]{%
5.75	3.084\\
5.75	4.02700000000001\\
6.25	4.02700000000001\\
6.25	3.084\\
5.75	3.084\\
};
\addplot [color=mycolorRand7, fill=mixedRandColor2, line width=0.4pt, solid, forget plot]
  table[row sep=crcr]{%
6.75	3.312375\\
6.75	3.884\\
7.25	3.884\\
7.25	3.312375\\
6.75	3.312375\\
};
\addplot [color=mycolorRand7, fill=mixedRandColor2, line width=0.4pt, solid, forget plot]
  table[row sep=crcr]{%
7.75	3.7295\\
7.75	4.2035\\
8.25	4.2035\\
8.25	3.7295\\
7.75	3.7295\\
};
\addplot [color=medianColor, line width = 0.3, forget plot]
  table[row sep=crcr]{%
0.75	1.279\\
1.25	1.279\\
};
\addplot [color=medianColor, forget plot]
  table[row sep=crcr]{%
1.75	2.6615\\
2.25	2.6615\\
};
\addplot [color=medianColor, forget plot]
  table[row sep=crcr]{%
2.75	3.46650000000001\\
3.25	3.46650000000001\\
};
\addplot [color=medianColor, forget plot]
  table[row sep=crcr]{%
3.75	2.002\\
4.25	2.002\\
};
\addplot [color=medianColor, forget plot]
  table[row sep=crcr]{%
4.75	2.5685\\
5.25	2.5685\\
};
\addplot [color=medianColor, forget plot]
  table[row sep=crcr]{%
5.75	3.5235\\
6.25	3.5235\\
};
\addplot [color=medianColor, forget plot]
  table[row sep=crcr]{%
6.75	3.57950000000002\\
7.25	3.57950000000002\\
};
\addplot [color=medianColor, forget plot]
  table[row sep=crcr]{%
7.75	3.95650000000001\\
8.25	3.95650000000001\\
};
\legend{\ours{}, \slr{m}, \mixr{m}};
\end{axis}
\end{tikzpicture}%
}
        \caption{}
        \label{fig:delayBox}
    \end{subfigure}
    \caption{Coexistence performance for the benchmark with semi-random \gls{urllc} devices. (a) Empirical CDF of URLLC devices' availability, $\hat{\alpha}_i^\Gamma$, where the shaded area around each plot indicates $99.99$\% confidence bounds. (b) Training delay, $d_k^{\mathrm{AI}}$, where each box plot represents the minimum, $25$th percentile, median, $75$th percentile, and maximum of the training delay distribution.}
    \label{fig:kpi}
    \vspace{-3mm}
\end{figure*}
\subsection{Semi-random URLLC Devices}
\figurename\,\ref{fig:kpi} shows the distribution of our main \glspl{kpi}. \figurename\,\ref{fig:availUrllc} illustrates the empirical \gls{cdf} of \gls{urllc} devices' availability, where each sample is the \gls{ul} or \gls{dl} availability of one \gls{urllc} device in one simulation during its whole simulation time. The shaded area around each plot represents the $99.99\%$ pointwise confidence bounds, determined using Greenwood’s formula~\cite{confBound}. \figurename\,\ref{fig:delayBox} depicts the training delay. Each box presents the minimum, $25$th percentile, median, $75$th percentile, and maximum values observed among the training delay samples. Together, these two figures present the coexistence performance.

In \figurename\,\ref{fig:availUrllc}, the availability distribution is identical for any arbitrary $m$ in \slr{m} because of the dedicated resource slices for each service.
As this figure shows, compared to \su{} and \slr{m}, the availability of the \gls{urllc} devices decreases in \mixr{m}, $\forall m{\in}\{15,20,40\}$, likely because of the introduced interference by \gls{ai} devices in the neighboring cells. Although the scheduler adjusts the \gls{mcs} to deal with this additional interference,\footnote{Note that such a decrease in availability occurs regardless of the scheduler configuration. For example, a higher target \gls{bler} cannot overcome the extra interference, and a lower target \gls{bler} leads to extra delay, both resulting in lower availability.} intensive distributed training traffic still affects the availability. Consequently, the $0.99$ requirement is met with a violation probability of around $0.1$, compared to roughly $0.01$ in \su{}, \slr{m}, and \ours{}, see Equation~\eqref{eq:availReq}.
In \mixr{m} baselines, most of the availability samples remain above or equal to $0.98$. However, unlike many conventional services, such a decrease is unacceptable for \gls{urllc} service.
Despite the impact of introducing the large load of the distributed learning service, we observe that our \ours{} solution keeps the \gls{urllc} devices' availability close to the \su{} and \slr{m} up to $\alpha^\mathrm{req}$.
When it comes to training delay performance, as \slr{m} and \mixr{m} boxes in \figurename\,\ref{fig:delayBox} and our results in~\cite{ganjInterplay} suggest, training delay increases with the number of selected devices (i.e., as $m$ grows). However, when $m{=}n$, it is more likely for the central node to wait excessively for stragglers and thus reach time-out (as in \slr{15} and \mixr{15}). Moreover, the lower training delay statistics in \mixr{15} and \mixr{20} than \slr{15} and \slr{20}, respectively, suggests that distributed learning service in \mixr{m} generally consumes more resources than the allocated resources in \slr{m}.
As this figure indicates, compared to the most competitive baseline (i.e., \mixr{15}), our \ours{} decreases the median training delay by $36\%$, while the maximum observed training delay is $2.6$\,s, which is $43\%$ less than that of \mixr{20}---the baseline with the lowest maximum observed training delay.
When \gls{urllc} traffic is well distributed and low per cell (as in this benchmark), \ours{} can easily satisfy the \gls{urllc} requirements, as baselines like \slr{m}. Yet, thanks to \ours{}, the low-priority service experiences a much better training delay than those baselines, as shown in \figurename\,\ref{fig:delayBox}. When \gls{urllc} traffic is high, meeting its requirements becomes more challenging for the baselines, but as we demonstrate in the next experiments, \ours{} still succeeds in managing the coexistence.
\begin{figure*}[!t]
\centering
    \begin{subfigure}[b]{0.49\textwidth}
        \centering {\footnotesize\pgfdeclarelayer{background}
\pgfdeclarelayer{foreground}
\pgfsetlayers{background,main,foreground}
\begin{tikzpicture}
\begin{pgfonlayer}{background}
\begin{axis}[%
width=72.0*0.99mm,
height=46.0mm,
at={(0.769in,0.482in)},
scale only axis,
xmin=14.5,
xmax=23.5,
xlabel style={font=\color{white!15!black}},
xlabel={Number of selected AI devices},
xtick = {15,16,17,18,19,20,21,22,23},
ymin=0,
ymax=0.65,
ytick = {0,0.1,0.2,0.3,0.4,0.5,0.6},
ylabel style={font=\color{white!15!black}},
ylabel={PMF},
axis background/.style={fill=white},
xmajorgrids,
ymajorgrids
]
\addplot[ycomb, color=ourSolColor, line width=1.2pt, mark=*, mark options={solid, fill=ourSolColor, ourSolColor}, forget plot] table[row sep=crcr] {%
15	0.587866666666667\\
16	0.201\\
17	0.138333333333333\\
18	0.0546\\
19	0.0130666666666667\\
20	0.00306666666666667\\
21	0.001\\
22	0.000733333333333333\\
23	3.333333333333337e-04\\
};
\coordinate (inset) at (axis description cs:0.974,0.93);
\end{axis}
\end{pgfonlayer}

\begin{pgfonlayer}{foreground}

\begin{semilogyaxis}[
    at={(inset)},
    anchor=north east,
    axis background/.style={fill=bgColor!10},
    width=52mm,
    height=42mm,
    log basis y = 10,
    ymax = 180,
    ymin = 1E-0,
    name=insetAxis,
    xtick = {19,20,21,22,23},
    xmin=18.75,
    xmax=23.25,
    yticklabels = { $10^{-4}$,$10^{-3}$, $10^{-2}$},
    grid style={dashed, dash pattern=on 4pt off 1pt on 2pt off 0.5pt},
    yminorgrids
]
\addplot[ycomb, color=ourSolColor, line width=1.2pt, mark=*, mark options={solid, fill=ourSolColor, ourSolColor}, forget plot] table[row sep=crcr] {%
19	130.666666666667\\
20	30.6666666666667\\
21	0010.0\\
22	0007.33333333333333\\
23	3.333333333333337\\
};
\end{semilogyaxis}
\end{pgfonlayer}

\begin{pgfonlayer}{main}
\end{pgfonlayer}

\end{tikzpicture}%
}
        \caption{}
        \label{fig:PMF}
    \end{subfigure}%
    \hfill
    \begin{subfigure}[b]{0.49\textwidth}
        \centering
        {\footnotesize
%
%
%
\begin{tikzpicture}

\begin{axis}[%
width=72.0*0.99mm,
height=46.0mm,
at={(0.758in,0.482in)},
scale only axis,
bar shift auto,
xmin=-0.2,
xmax=51.2,
xlabel style={font=\color{white!15!black}},
xlabel={AI device number},
xtick = {1,5,10,15,20,25,30,35,40,45,50},
ymin=0,
ymax=1.008,
ylabel style={font=\color{white!15!black}},
ylabel={Device selection ratio},
ytick={0,0.2,0.4,0.6,0.8,1},
yticklabels={$0$,$0.2$,$0.4$,$0.6$,$0.8$,$1$},
axis background/.style={fill=white},
xmajorgrids,
ymajorgrids
]
\addplot[ybar, bar width=0.8, fill=ourSolColor, draw=black, area legend,postaction={pattern= crosshatch dots, pattern color = black, dash pattern=on \pgflinewidth off 0.75pt}] table[row sep=crcr] {%
1	0\\
2	0\\
3	0.04\\
4	0.64\\
5	1\\
6	0\\
7	0\\
8	0\\
9	0.4\\
10	0.92\\
11	0.78\\
12	0\\
13	0.86\\
14	0.12\\
15	0.18\\
16	0\\
17	0\\
18	0.06\\
19	0\\
20	0\\
21	0.02\\
22	0\\
23	0\\
24	0\\
25	0\\
26	0\\
27	0.94\\
28	0\\
29	0\\
30	0.86\\
31	0.98\\
32	0\\
33	0.96\\
34	0.14\\
35	0\\
36	1\\
37	0\\
38	0\\
39	0\\
40	0\\
41	0.02\\
42	0.54\\
43	1\\
44	0\\
45	0.9\\
46	0.94\\
47	0\\
48	0.88\\
49	0.86\\
50	0.98\\
};
\end{axis}
\end{tikzpicture}%
}
        \caption{}
        \label{fig:selRatio}
    \end{subfigure}
    \caption{Device selection policy of \ours{} for the benchmark with semi-random \gls{urllc} devices. (a) Empirical PMF of the number of selected devices, $m_k$. (b) Participation ratio of each device.}
    \label{fig:dRlAgentPerf}
    \vspace{-2mm}
\end{figure*}

\figurename\,\ref{fig:dRlAgentPerf} demonstrates the \ours{}'s device selection policy in the evaluation phase, $\bm{\pi}_k^\mathrm{u}$. \figurename\,\ref{fig:PMF} shows the empirical \gls{pmf} of the number of selected \gls{ai} devices, $m_k$. As this figure indicates, \ours{} selected at least $1$ extra backup \gls{ai} device (i.e., $m_k{-}n\geq 1$) for more than $40\%$ of iterations. Such selection of extra backup devices implies that our device selection solution could still leverage the diversity introduced by extra backup \gls{ai} devices, even in our bandwidth-limited scenario. Carefully selected backup devices reduce the sensitivity to the straggler problem and, therefore, reduce the overall latency without substantially impacting the interference footprint. 
\figurename\,\ref{fig:selRatio} represents the participation ratio of each \gls{ai} device. As it shows, $9$ \gls{ai} devices are selected in over $90\%$ of the iterations. Furthermore, $24\%$ of the \gls{ai} devices have a selection ratio ranging between $0.1$ and $0.9$, depending on their channel conditions, the loads of their corresponding cells, and the deployment specifics. 
In extreme cases, some of the devices may always (or never) be selected to participate in the distributed training.

\subsection{Random URLLC Devices}
\figurename\,\ref{fig:kpiRand} illustrates the empirical \gls{cdf} of availability for \gls{urllc} devices (in \figurename\,\ref{fig:availUrllcRand}) and training delay of distributed learning service (in \figurename\,\ref{fig:delayBoxRand}). As \figurename\,\ref{fig:availUrllcRand} shows, up to $\alpha^{\mathrm{req}}$, \ours{} keeps the availability of \gls{urllc} devices close to \su{} (by only $0.0007$ increase in $\gamma$), even though the \gls{urllc} traffic appears at random locations in different seeds.
Surprisingly, \slo{} meets a $0.99$ availability requirement at a violation probability of $\gamma{=}0.019$, which is around $2.5$x greater than \su{} and \ours{}, and even $14\%$ more than \mixo{}. It is worth noting that the availability distributions for \slo{} and \slr{m} are identical due to the consistent operation of the \gls{urllc} slice in both baselines.
In this benchmark, with twice the \gls{urllc} devices than the semi-random benchmark, many devices may be associated with the same \gls{gnb}. Thus, in \slo{}, the dedicated slice may be insufficient for certain \glspl{gnb} across different seeds, leading to lower availability performance than baselines in which services share wireless resources. Nevertheless, as \figurename\,\ref{fig:delayBoxRand} shows, \slo{} reaches higher training delays than other baselines too (with the median of $1.8$\,s and even occasional time-outs), indicating that the static slicing might cause resource deprivation for both services. However, 
Since \mixo{} only considers the distributed learning service performance, its $\gamma$ to support $\alpha^{\mathrm{req}}$ is $2.24$x and $2.05$x the violation probabilities of \su{} and \ours{}, respectively. In \figurename\,\ref{fig:delayBoxRand}, \ours{} achieves $1.42$\,s median training delay, which is $34\%$ higher than \mixo{}. This suggests that \ours{} compensates an additional $360$\,ms on the median training delay of the low-priority service to halve the violation probability of \gls{urllc} service's availability requirement.


\begin{figure*}[!t]
    \centering
    \begin{subfigure}[b]{1\columnwidth}
        \centering 
        {\footnotesize\input{./Components/Figs/journal/availCdfRandV3}}
        \caption{}
        \label{fig:availUrllcRand}
    \end{subfigure}%
    \hfill\hfill
    \begin{subfigure}[b]{1\columnwidth}
        \centering
        {\footnotesize
%
%
\definecolor{mycolorRand7}{rgb}{0.12, 0.3, 0.17}
\definecolor{medianColor}{rgb}{0.7, 0.11, 0.11}

\definecolor{mycolorSlic}{rgb}{0.45, 0.31, 0.59}
\definecolor{mycolorSlic}{rgb}{0.5, 0.0, 0.5}
\definecolor{mycolorSlic}{rgb}{0.47, 0.32, 0.66}

\definecolor{mycolorDrl}{rgb}{0.0, 0.0, 0.55}
\definecolor{mycolorDrl}{rgb}{0.0, 0.14, 0.4}
\definecolor{mycolorDrl}{rgb}{0.25, 0.41, 0.88}
\begin{tikzpicture}

\begin{axis}[%
width=72.0*0.99mm,
height=46.0mm,
at={(0.769in,0.482in)},
scale only axis,
unbounded coords=jump,
xmin=0.5,
xmax=3.5,
xticklabels={,,}
ymin=0.8,
ymax=4*1.0125,
ylabel style={font=\color{white!15!black}},
ylabel={Training delay (s)},
axis background/.style={fill=white},
xmajorgrids,
ymajorgrids,
legend style={at={(7.7cm,-0.8cm)}, anchor=north east,legend columns=-1, fill=white, inner sep=0.8pt, outer sep=0.01pt, row sep=0.01pt},
]
\addplot [color=mycolorDrl,line width=0.75pt, solid, forget plot]
  table[row sep=crcr]{%
1	1.5525\\
1	3.99299999999999\\
};  
\addplot [color=mycolorSlic, line width=0.75pt, solid, forget plot]
  table[row sep=crcr]{%
2	1.894\\
2	10\\
};
\addplot [color=mycolorOnlyAiOpt, line width=0.75pt, solid, forget plot]
  table[row sep=crcr]{%
3	1.1207\\
3	1.9025\\
};
\addplot [color=mycolorRand7, line width=0.75pt, solid, forget plot]
  table[row sep=crcr]{%
4	2.2455\\
4	10\\
};
\addplot [color=mycolorRand7, line width=0.75pt, solid, forget plot]
  table[row sep=crcr]{%
5	2.8905\\
5	4.705\\
};

\addplot [color=mycolorDrl, line width=0.75pt, solid,  forget plot]
  table[row sep=crcr]{%
1	1.0065\\
1	1.32125\\
};
\addplot [color=mycolorSlic, line width=0.75pt, solid, forget plot]
  table[row sep=crcr]{%
2	1.743\\
2	1.418\\
};
\addplot [color=mycolorOnlyAiOpt, line width=0.75pt, solid, forget plot]
  table[row sep=crcr]{%
3	1.0165\\
3	0.9185\\
};
\addplot [color=mycolorRand7, fill=mixedRandColor2, line width=0.75pt, solid, forget plot]
  table[row sep=crcr]{%
4	1.1975\\
4	1.834\\
};
\addplot [color=mycolorRand7, fill=mixedRandColor2, line width=0.75pt, solid, forget plot]
  table[row sep=crcr]{%
5	1.3595\\
5	2.3355\\
};

\addplot [color=mycolorDrl, line width=0.75pt, solid, forget plot]
  table[row sep=crcr]{%
0.875	3.99299999999999\\
1.125	3.99299999999999\\
};
\addplot [color=mycolorSlic, line width=0.75pt, solid,  forget plot]
  table[row sep=crcr]{%
1.875	10\\
2.125	10\\
};
\addplot [color=mycolorOnlyAiOpt, line width=0.75pt, solid, forget plot]
  table[row sep=crcr]{%
2.875	1.9025\\
3.125	1.9025\\
};
\addplot [color=mycolorRand7, fill=mixedRandColor2, line width=0.75pt, solid, forget plot]
  table[row sep=crcr]{%
3.875	10\\
4.125	10\\
};
\addplot [color=mycolorRand7, fill=mixedRandColor2, line width=0.75pt, solid, forget plot]
  table[row sep=crcr]{%
4.875	4.705\\
5.125	4.705\\
};

\addplot [color=mycolorDrl, line width=0.75pt, solid, forget plot]
  table[row sep=crcr]{%
0.875	1.0065\\
1.125	1.0065\\
};
\addplot [color=mycolorSlic, line width=0.75pt, solid,  forget plot]
  table[row sep=crcr]{%
1.875	1.418\\
2.125	1.418\\
};
\addplot [color=mycolorOnlyAiOpt, line width=0.75pt, solid, forget plot]
  table[row sep=crcr]{%
2.875	0.9185\\
3.125	0.9185\\
};
\addplot [color=mycolorRand7, fill=mixedRandColor2, line width=0.75pt, solid, forget plot]
  table[row sep=crcr]{%
3.875	1.1975\\
4.125	1.1975\\
};
\addplot [color=mycolorRand7, fill=mixedRandColor2, line width=0.75pt, solid, forget plot]
  table[row sep=crcr]{%
4.875	1.3595\\
5.125	1.3595\\
};

\addplot [color=mycolorDrl, fill=ourSolColor, line width=.25pt, solid, area legend]
  table[row sep=crcr]{%
0.75	1.32125\\
0.75	1.5525\\
1.25	1.5525\\
1.25	1.32125\\
0.75	1.32125\\
};

\addplot [color=mycolorSlic, fill=slicingColor, line width=0.4pt, area legend]
  table[row sep=crcr]{%
1.75	1.743\\
1.75	1.894\\
2.25	1.894\\
2.25	1.743\\
1.75	1.743\\
};

\addplot [color=mycolorOnlyAiOpt, fill=mixedRandColor2, line width=0.4pt, area legend]
  table[row sep=crcr]{%
2.75	1.0165\\
2.75	1.1207\\
3.25	1.1207\\
3.25	1.0165\\
2.75	1.0165\\
};

\addplot [color=mycolorRand7, fill=mixedRandColor2, line width=0.4pt, solid, area legend]
  table[row sep=crcr]{%
3.75	1.834\\
3.75	2.2455\\
4.25	2.2455\\
4.25	1.834\\
3.75	1.834\\
};
\addplot [color=mycolorRand7, fill=mixedRandColor2, line width=0.4pt, solid, area legend]
  table[row sep=crcr]{%
4.75	2.3355\\
4.75	2.8905\\
5.25	2.8905\\
5.25	2.3355\\
4.75	2.3355\\
};

\addplot [color=medianColor, line width = 0.3, forget plot]
  table[row sep=crcr]{%
0.75	1.4225\\
1.25	1.4225\\
};
\addplot [color=medianColor, forget plot]
  table[row sep=crcr]{%
1.75	1.796\\
2.25	1.796\\
};
\addplot [color=medianColor, forget plot]
  table[row sep=crcr]{%
2.75	1.0575\\
3.25	1.0575\\
};
\addplot [color=medianColor, forget plot]
  table[row sep=crcr]{%
3.75	2.0325\\
4.25	2.0325\\
};
\addplot [color=medianColor, forget plot]
  table[row sep=crcr]{%
4.75	2.594\\
5.25	2.594\\
};
\legend{\ours{}, \slo{}, \mixo{}};

\node[anchor=west] (source) at (axis cs:2,4.15){};
\node (destination) at (axis cs:1.56,3.2){Maximum $=10$\,s};
\draw[-stealth, dash pattern=on \pgflinewidth off 0.8pt,line width=0.6pt](source)--(destination);
\end{axis}
\end{tikzpicture}%
}
        \caption{}
        \label{fig:delayBoxRand}
    \end{subfigure}
    \caption{Coexistence performance for the benchmark with random \gls{urllc} devices. (a) Empirical CDF of URLLC devices' availability, $\hat{\alpha}_i^\Gamma$, where the shaded area around each plot indicates $99.99$\% confidence bounds. (b) Training delay, $d_k^{\mathrm{AI}}$, where each box plot represents the minimum, $25$th percentile, median, $75$th percentile, and maximum of the training delay distribution.}
    \label{fig:kpiRand}
    \vspace{-2mm}
\end{figure*}

\section{Conclusions}\label{sec:conculsions}
In this paper, we investigated the performance optimization of distributed learning services coexisting with \gls{urllc} services. We proposed \gls{bsac}-CoEx, a novel framework that utilizes a carefully selected subset of devices for distributed training, aiming to minimize the convergence time of distributed learning while maintaining \gls{urllc}'s stringent availability requirements.
Our comprehensive \gls{3gpp}-compliant simulations indicate that our scheme can significantly reduce the total training delay while maintaining the availability of \gls{urllc} devices close to that in the single service scenario (i.e., when \gls{urllc} devices exclusively utilize all network resources). 
Our results provide valuable insights into adaptively controlling distributed training traffic through device selection, ensuring sustainable coexistence between these two services.

An alternative approach to controlling the distributed training load is through the quantization of exchanged messages, thereby reducing the communication overhead per iteration and, consequently, the interference footprint. However, this reduction comes at the expense of requiring additional iterations to reach convergence. Potential future work includes developing novel methods that dynamically adjust the quantization level based not only on the distributed learning algorithm but also on the \gls{urllc} load and interference footprint on the network.


\ifthenelse{\boolean{appendixversion}}{
  \renewcommand{\theequation}{A.\arabic{equation}}
\setcounter{equation}{0}
{\appendices

\section{Deriving the Number of Iterations to Converge}\label{app:a}
The first-order necessary condition for local optimality is given by the vanishing of gradients. However, depending on the characteristics of the optimization landscape, equivalent conditions may be defined based on the objective functions. In the following subsections, we explore different optimization scenarios and provide examples of the bounds that can be leveraged to engineer $K_{\min}$. These alternative theoretical approaches complement the numerical design of $K_{\min}$, as illustrated in Equation~\eqref{eq:Kmin}.

Towards solving the distributed optimization problem \eqref{eq: MainOptimProblem}, the central node performs the gradient descent update at the $k$th iteration as
\begin{equation}\label{eq:algUpdate}
\bw_{k+1} = \bw_k - \frac{\eta_k}{n}\sum_{i\in\calN_{n,k}}\!\!\widehat{\nabla}f_i{(\bw_k)},
\end{equation}
where the right hand side of the equation represents $A{(\cdot)}$ in \eqref{eq:globalUp}, $\widehat{\nabla}f_i{(\bw_k)}$ is the true gradient's noisy estimation at the $i$th \gls{ai} device, and $\calN_{n,k}$ is the set of \gls{ai} devices from which central node received the first $n$ local updates at the $k$th iteration. We assume that each \gls{ai} device employs mini-batch gradient descent, and to simplify our notation, the size of the mini-batches is assumed to be the same for all devices in all iterations. For the sake of simplicity, we assume that the gradients' noise is \gls{iid}~\cite{gradientNoise}. Notice that the numerical approximations of $K_{\min}$ do not rely on the gradient noise assumption, and we need this assumption only for the theoretical approximation of $K_{\min}$.

The overall gradient estimate using the local estimates of the \gls{ai} devices can be obtained as
\begin{equation}\label{eq:noisyGradient}
\frac{1}{n}\!\sum_{i\in\calN_{n,k}}\!\!\widehat{\nabla}f_i{(\bw_k)} \coloneq \nabla f{(\bw_k)}+\Bar{\bolde}_k^{(n)},
\end{equation}
where $\Bar{\bolde}_k^{(n)} \coloneq \frac{1}{n}\!\sum_{i\in\calN_{n,k}}\!\!\bolde_{i,k}$, and $\bolde_{i,k}$ is the residual term of the $i$th device's estimate at the $k$th iteration, while $\nabla f{(\bw_k)}$ is the true gradient (i.e., the gradient of the batch gradient descent on centralized training).

\subsection{DGD, Strongly Convex Objective Functions} \label{app:1}
\begin{assumption}\label{as:1}
\cite{Bottou2018SIAM,flConvergNoniid} The objective functions $f_i$, $\forall i\in\calN$, are all $L$-smooth, with Lipschitz constant $L>0$.
\end{assumption}

\begin{assumption}\label{as:2}
\cite{Bottou2018SIAM,flConvergNoniid} The objective functions $f_i$, $\forall i\in\calN$, are all strongly convex, with constant $\mu>0$.
\end{assumption}

\begin{assumption}\label{as:3}
\cite{Bottou2018SIAM} There exist $\beta_2\geq (\beta_1+1)^2 > 0$ that, for $\forall k\in \left[K\right]$ and $\forall i\in \calN$, the objective function $f(\bw)$ and the \gls{dgd} algorithm have the following limits:
\begin{subequations}\label{eq:momentLimits}
\begin{align}
    & \nabla f{(\bw_k)}^\intercal \mathbb{E}{\left[\bolde_{i,k}\right]}\geq \beta_1 \norm{\nabla f{(\bw_k)}}_2^2 , \label{eq:subspace}\\
    & \nabla f{(\bw_k)}^\intercal \mathbb{E}{\left[\bolde_{i,k}\right]} \leq \beta_2 \norm{\nabla f{(\bw_k)}}_2^2. \label{eq:upperBoundLim}
\end{align}
\end{subequations}
It is worth noting that \eqref{eq:subspace} implies that the noisy estimation of the gradient is on the same half space with the true gradient, and \eqref{eq:upperBoundLim} is a weaker assumption of the bounded variance of $\sum_{i\in\calN_{n,k}}\!\!\widehat{\nabla}f_i{(\bw_k)}/n$, and only bounds it by the actual gradient, $\nabla f(\bw_k)$.
\end{assumption}
\begin{assumption}\label{as:4}
\cite{Bottou2018SIAM,flConvergNoniid} The variance of the gradient norms in each device is bounded, i.e.,
\begin{equation}\label{eq:sigma}
    \mathbb{E}{\left[\norm{\bolde_{i,k}}_2^2\right]}\leq \sigma^2, \forall k\in \left[K\right], \forall i\in \calN.
\end{equation}
Since, $\Bar{\bolde}_k^{(n)}$ is an unbiased estimator of $\mathbb{E}{\left[\bolde_{i,k}\right]}$, we have
\begin{equation}
    \mathbb{E}{\left[\norm{\Bar{\bolde}_k^{(n)}}_2^2\right]}\leq \frac{\sigma^2}{n}, \forall k\in \left[K\right].
\end{equation}
Notably, from \eqref{eq:upperBoundLim} and \eqref{eq:sigma}, we can derive 
\begin{equation}\label{eq:sgdVarComb}
    \mathbb{E}{\left[\norm{\widehat{\nabla}f_i{(\bw_k)}}_2^2\right]}\leq\sigma^2+\beta_2 \norm{\nabla f{(\bw_k)}}_2^2.
\end{equation}
\end{assumption}
Then, if Assumptions~\ref{as:1}-\ref{as:4} hold, for a fixed learning rate $\eta$ that is $0<\eta\leq\frac{\beta_1+1}{(2\beta_2+1)L}$, we have \cite[Theorem 4.6]{Bottou2018SIAM}
\begin{multline}\label{eq:convexConverge}
\mathbb{E}{\left[f{\left(\bw_k\right)} {-} f{\left(\bw^\star\right)}\right]} \leq \frac{\eta L \sigma^2}{2n\beta_1 \mu} +\\ 
{\left(1{-}\eta \beta_1 \mu\right)^{k-1}}{\left(f{\left(\bw_1\right)}{-}f{\left(\bw^\star\right)}{-}\frac{\eta L\sigma^2}{2n\beta_1 \mu}\right)},
\end{multline}
where the first term represents the gap to the expected optimal value that \gls{dgd} converges to when $k{\to} \infty$ for a fixed learning rate, and the second term is the convergence rate. Using the learning rate bound and Assumption \ref{as:3}, and the fact that $\mu {\leq} L$(as a result of Assumption \ref{as:1} and \ref{as:2}), we can derive that $\eta \beta_1 \mu < 1$, and hence, $(1-\eta \beta_1 \mu)$ is a contraction factor.

Let us assume that our initial point is within a bounded region with respect to the final point that we can converge (i.e., the last parenthesis in \eqref{eq:convexConverge} is less than or equal to $W^{\mathrm{A}}$). Note that the additional term of $\eta L \sigma^2/{2n\beta_1 \mu}$ reflects that \gls{dgd} cannot converge to the optimal value, but instead, to a neighborhood of $f(\bw^\star)$.
Then, the minimum required number of iterations, $K_{\min}$, to reach $\epsilon$-optimality becomes
\begin{align}\label{eq:kConvex}
K_{\min} \geq \log_{{\left(1{-}\eta \beta_1 \mu\right)}}&{\left(\epsilon{-}\frac{\eta L\sigma^2}{2n\beta_1 \mu}\right)} - \log_{{\left(1{-}\eta \beta_1 \mu\right)}}\!\!{\left({W}\right)} +1.
\end{align}
Then, \eqref{eq:kConvex} can be simplified as
\begin{equation}\label{eq:kConvexSimpleApp}
K_{\min} \geq \log_{b}{\left(\frac{W^{\mathrm{A}}}{\epsilon{-}\frac{z^{\mathrm{A}}}{n}}\right)} +1,
\end{equation}
where $b\coloneq1/\left(1{-}\eta \beta_1 \mu\right)>1$, and $z^{\mathrm{A}}$ is a positive constant which depends on the learning rate, Lipschitz constant, strong convexity, and the error in the gradient estimates for $n{=}1$. 

\subsection{DGD, Non-convex Objective Functions}\label{app:2}
\begin{assumption}\label{as:5}
\cite{Bottou2018SIAM} The objective functions $f_i$, $\forall i \in \calN$, are lower bounded by a scalar $f_{\mathrm{inf}}$ for all sequences of $\bw_k$.
\end{assumption}
The non-convex objective functions may contain several local minima and other stationary points. Therefore, we define the convergence criteria on the gradient.
Then, if Assumptions~\ref{as:1}, \ref{as:3}-\ref{as:5} hold, for a fixed learning rate satisfying $0<\eta\leq \frac{\beta_1+1}{L(2\beta_2+1)}$, we have \cite[Theorem 4.8]{Bottou2018SIAM}
\begin{equation}\label{eq:nonConvexIneq}
\mathbb{E}{\left[\frac{1}{K}\!\sum_{k=1}^K \norm{\nabla f{(\bw_k)}}_2^2\right]} \leq \frac{\eta L \sigma^2}{n{(\beta_1{+}1)}} {+} \frac{2\left(f{(\bw_1)}{-}f_{\mathrm{inf}}\right)}{\eta{(\beta_1{+}1)}K}.
\end{equation}
To understand \eqref{eq:nonConvexIneq}, consider centralized training and batch gradient descent, where there exist no gradient noise, and $\sigma^2$ becomes zero, resulting in $\norm{\nabla f{(\bw_k)}}_2\to0$ as $K$ enlarges. However, in \gls{dgd}, the average norm of gradients converges to ${\eta L \sigma^2}/{n(\beta_1{+}1)}$.
Now, the required number of iterations, $K_{\min}$, to reach $\epsilon$-optimality becomes
\begin{equation}\label{eq:nonConvexKmin}
K_{\min}\geq \frac{2\left(f{(\bw_1)}-f_{\mathrm{inf}}\right)}{\eta{(\beta_1{+}1)}\left(\epsilon{-}\frac{\eta L \sigma^2}{n\left(\beta_1{+}1\right)}\right)}.
\end{equation}
In \eqref{eq:nonConvexKmin}, we observe that loosening the convergence criteria (i.e., as $\epsilon$ increases) reduces the required number of iterations to reach $\epsilon$-optimality gap. We can simplify \eqref{eq:nonConvexKmin} as
\begin{equation}\label{eq:kNonConvexSimpleApp}
K_{\min}\geq \frac{W^\mathrm{B}}{\epsilon{-}\frac{z^\mathrm{B}}{n}},
\end{equation}
where $W^\mathrm{B} {\coloneq} 2\left(f{(\bw_1)}{-}f_{\mathrm{inf}}\right)/\eta{(\beta_1{+}1)}$, and $z^\mathrm{B}$ is a function of the learning rate, Lipschitz constant, and error in the gradient estimates when $n{=}1$. Note that $\epsilon$ should be set to a value larger than the neighborhood \gls{dgd} can potentially converge to.

\subsection{FL, Strongly Convex Objective Functions}\label{app:3}
There are two main differences between \gls{fl} and \gls{dgd}, 
i) there could be several local iterations in each \gls{ai} device between two communications, 
and ii) the model parameters (i.e., the weights of the \glspl{dnn}) are communicated, rather than the gradients in \gls{dgd}. Hence, on the local update, each \gls{ai} device performs \eqref{eq:algUpdate} for $E$ times before updating the global iteration $k$, as \eqref{eq:localUp}. 

\begin{assumption}\label{as:6}
\cite{flConvergNoniid} The variance of the gradient estimates in each \gls{ai} device is bounded, i.e.,
\begin{equation}\label{eq:boundedGradient}
    \mathbb{E}{\left[\norm{\widehat{\nabla}f_i{(\bw_k)}}_2^2\right]}\leq G^2.
\end{equation}
Note that \eqref{eq:boundedGradient} is an stricter assumption than Assumptions~\ref{as:3} and \ref{as:4}, combined, as shown in \eqref{eq:sgdVarComb}.
\end{assumption}

If Assumptions~\ref{as:1}, \ref{as:2}, \ref{as:4}, and \ref{as:6} hold, and $n$ \gls{ai} devices are selected uniformly at each iteration, then for a diminishing learning rate $\eta_k = 2/\mu{(\xi+k+\kappa)}$, where $\kappa({\in} {[E{-}1]})$ is the local iteration number and $\xi \coloneq \max\left\{8L/\mu,E\right\}$, the following inequality holds \cite[Theorem 3]{flConvergNoniid}:
\begin{equation}\label{eq:flConvergence}
\mathbb{E}{\left[f{\left(\bw_k\right)} {-} f{\left(\bw^\star\right)}\right]}\leq \frac{2L{\left(\frac{\sigma^2}{N}{+}8{(E{-}1)^2}{+} \rho E^2G^2{+}\xi G^2\right)}}{\mu^2{(\xi{+}k{+}\kappa{-}1)}},
\end{equation}
where $\rho\coloneq\frac{4{(N-n)}}{n{(N-1)}}$. Hence, the minimum number of global iterations (i.e., rounds of communications) to attain $\epsilon$-optimality approximately becomes~\cite{flConvergNoniid}
\begin{equation}\label{eq:flEpsAccuracyApp}
K_{\min} \propto \frac{1}{\epsilon}\left[\left(1+\frac{1}{n}\right)EG^2 + \frac{\frac{\sigma^2}{N}+G^2}{E}+G^2\right],
\end{equation}
where we assumed $\xi=\calO{(1+E)}$.

\section{Base 3GPP Channel Model}\label{app:channelModel}
To model the channel, we consider a \gls{mimo} system in which we leverage the time-varying 3D spatial channel model from \gls{3gpp} in~\cite{3GPP38901}. In this model, channels are characterized via clustering the multipath components, arriving at antenna arrays, in delay and double-directional angle (i.e., the zenith and azimuth of the \glspl{aoa} at the receiver and \glspl{aod} at the transmitter). 
For simplicity, let us assume that $N_\mathrm{g}$ and $N_\mathrm{d}$ are respectively the number of antenna elements of \gls{gnb} and devices.
We denote by $\bm{H}_{x,y}(\tau;t) \in \mathds{C}^{N_\mathrm{d}\times N_\mathrm{g}}$ the baseband channel response at time $t$ to an input impulse at time $t-\tau$, between the $x$th device and $y$th \gls{gnb} in \gls{dl}.
Then, an entry of $\bm{H}_{x,y}(\tau;t)$ for the $p$th receiving antenna element and $q$th transmitting antenna element can be computed as
\begin{multline}\label{eq:channel}
\left[\bm{H}_{x,y}{\left(\tau;t\right)}\right]_{p,q} {\coloneqq} \sqrt{\beta^{x,y}}\sum_{l=1}^{N_\mathrm{c}} \sqrt{P_{l}^{x,y}} \sum_{s=1}^{N_\mathrm{s}} {\left(\bm{g}_{p}^{x,y}{\left(t,l,s\right)}\right)}^\intercal\\ \bm{F}_{\mathrm{xp}}^{x,y}{\left(t,l,s\right)} 
\bm{g}_{q}^{x,y}{\left(t,l,s\right)}e^{j\Upsilon_{p,q}^{x,y}{\left(t,l,s\right)}}\delta{\left(\tau-\tau_{\mathrm{p},l,s}\right)},
\end{multline}
where $N_{\mathrm{c}}$ and $N_{\mathrm{s}}$ are, respectively, the number of clusters and rays. $\beta^{x,y}$ accounts for path loss and shadowing, which vary depending on the scenario (e.g., urban, indoor, etc.), and $P_{l}^{x,y}$ is a function of $l$th cluster normalized power. In Appendix~\ref{app:inf}, we have specified details of $\beta^{x,y}$ for our \gls{inf-dh} scenario.
Besides, ${\bm{g}_{p}^{x,y}\!{\left(\cdot\right)}}$ is the field patterns of the $p$th receiving element that the $s$th ray of the $l$th cluster has in the direction defined by the arriving zenith and azimuth angles, $\bm{F}_{\mathrm{xp}}^{x,y}\!{\left(\cdot\right)}$ is a $2{\times}2$ matrix modeling the cross polarization power ratio for the $s$th ray of the $l$th cluster, $\bm{g}_{q}^{x,y}\!{\left(\cdot\right)}$ is the field patterns of the $q$th transmitting element that the $s$th ray of the $l$th cluster has in the direction defined by the departing zenith and azimuth angles, $\Upsilon_{p,q}^{x,y}{\left(\cdot\right)}$ is a function of the location vector of the $p$th receiving and $q$th transmitting element as well as the Doppler frequency, and finally, $\tau_{\mathrm{p},l,s}$ is the propagation delay of the $s$th ray in the $l$th cluster. For \gls{ul}, $\bm{H}_{x,y}{\left(\tau;t\right)}$ can be derived by swapping $p$ and $q$ in~\eqref{eq:channel}.

\section{InF-DH Scenario of the Base 3GPP Channel Model}\label{app:inf}
In this appendix, we specify the parameters of the base 3GPP channel model based on \gls{inf-dh} scenario~\cite{3GPP38901}, used in our simulations.
The clutters in \gls{inf-dh} scenario typically represent small to medium-sized metallic machines and irregularly shaped objects. To calculate $\beta_{l}^{x,y}$ in \eqref{eq:channel}, the path loss is calculated by tracing the degradation in signal strength over distance under \gls{los} and \gls{nlos} circumstances. The path loss under \gls{los} and \gls{nlos} assumptions are given by~\cite{3GPP38901}
\begin{subequations}\label{eq:pathLoss}
\begin{align}
    &PL_{\mathrm{LOS}}{(d_{x,y}^{\mathrm{3D}})}\mathrm{[dB]}{=}31.84{+}21.5\log_{10}\!{\left(d_{x,y}^{\mathrm{3D}}\right)}{+}19\log_{10}\!{\left(f_{\mathrm{c}}\right)}, \label{eq:PL_LOS}\\
    &PL_{\mathrm{NLOS}}{(d_{x,y}^{\mathrm{3D}})} \mathrm{[dB]} {=} \max{\left(PL_{\mathrm{LOS}}{(d_{x,y}^{\mathrm{3D}})}, PL_{\mathrm{DH}}{(d_{x,y}^{\mathrm{3D}})} \right)},\label{eq:PL_NLOS}
\end{align}
\end{subequations}
where
\begin{equation}\label{eq:PL_InFDH}
PL_{\mathrm{DH}}{(d_{x,y}^{\mathrm{3D}})} \mathrm{[dB]}{=}33.63+21.9\log_{10}{\left(d_{x,y}^{\mathrm{3D}}\right)}+20\log_{10}{\left(f_{\mathrm{c}}\right)}.
\end{equation}
In above equations, $d_{\mathrm{3D}}$ and $f_{\mathrm{c}}$ denote the $3$D distance between the device and \gls{gnb}, and the center frequency, respectively. 
In \gls{inf-dh}, the \gls{los} probability is described by~\cite{3GPP38901}
\begin{equation}\label{eq:Pr_LOS}
\mathrm{Pr}_{\mathrm{LOS}}{(d_{x,y}^{\mathrm{2D}})} {=} \exp\left(\frac{d_{x,y}^{\mathrm{2D}}\ln\!{\left(1-r_\mathrm{clut}\right)}\left(h_\mathrm{clut}-h_\mathrm{device}\right)}{d_{\mathrm{clut}}\left(h_\mathrm{gNB}-h_\mathrm{device}\right)}\right),
\end{equation}
where $d_{\mathrm{2D}}$ represents the ground distance between \gls{gnb} and the device. Besides, $h_\mathrm{gNB}$, $h_\mathrm{device}$, $d_{\mathrm{clut}}$, $h_{\mathrm{clut}}$, and $r_{\mathrm{clut}}$ denote the \gls{gnb}'s antenna height, devices' antenna height, the typical clutter size, height and density, and are set in our simulations to $8$\,m, $1.5$\,m, $2$\,m, $6$\,m, and $60\%$, respectively. The shadowing for \gls{los} and \gls{nlos} is assumed to follow a zero-mean log-normal distribution with standard deviation $4.3$ and $4$ in dB, respectively. In our link level simulations, we first set the position of the $4$ \glspl{gnb}, as shown in \figurename\,\ref{fig:simulationSetup}. Then, for each pair of possible device positions and the $4$ \gls{gnb} positions, we generate uncorrelated link conditions with $\mathrm{Pr}_{\mathrm{LOS}}{(d_{\mathrm{2D}})}$ for \gls{los}, and ${1-\mathrm{Pr}_{\mathrm{LOS}}{(d_{\mathrm{2D}})}}$ for \gls{nlos}. Assuming that $\Omega_{\mathrm{LOS}}$ and $\Omega_{\mathrm{NLOS}}$ are random variables denoting the shadowing for \gls{los} and \gls{nlos}, respectively, then we have
\begin{equation} \label{eq:beta}
  \beta^{x,y} \coloneq\begin{cases}
    10^{\frac{-PL_{\mathrm{LOS}}{\left(d_{x,y}^{\mathrm{3D}}\right)-\Omega_{\mathrm{LOS}}}}{10}}, & \text{with probability $\mathrm{Pr}_{\mathrm{LOS}}$},\\
    10^{\frac{-PL_{\mathrm{NLOS}}{(d_{x,y}^{\mathrm{3D}})-\Omega_{\mathrm{NLOS}}}}{10}}, & \text{with probability $1-\mathrm{Pr}_{\mathrm{LOS}}$}.
  \end{cases}
\end{equation}
Nevertheless, the large-scale parameters are generated with a correlation distance of $10$\,m in the horizontal plane. Then, we followed the spatial consistency procedure in \cite[\S7.5,\S7.6.3]{3GPP38901} to generate small-scale parameters and channel coefficients and used the parameters in \cite[Table 7.5-6 Part-3]{3GPP38901}.
}
}{
}

\ifCLASSOPTIONcaptionsoff
\fi

%
%
\Urlmuskip=0mu plus 1mu\relax
\end{NoHyper}
\bibliographystyle{IEEEtran}
\vspace{-2mm}
\bibliography{Components/Metafiles/IEEEabrv.bib,Components/Metafiles/ref.bib}{}

\begin{thebibliography}{10}
\providecommand{\url}[1]{#1}
\csname url@samestyle\endcsname
\providecommand{\newblock}{\relax}
\providecommand{\bibinfo}[2]{#2}
\providecommand{\BIBentrySTDinterwordspacing}{\spaceskip=0pt\relax}
\providecommand{\BIBentryALTinterwordstretchfactor}{4}
\providecommand{\BIBentryALTinterwordspacing}{\spaceskip=\fontdimen2\font plus
\BIBentryALTinterwordstretchfactor\fontdimen3\font minus
  \fontdimen4\font\relax}
\providecommand{\BIBforeignlanguage}[2]{{%
\expandafter\ifx\csname l@#1\endcsname\relax
\typeout{** WARNING: IEEEtran.bst: No hyphenation pattern has been}%
\typeout{** loaded for the language `#1'. Using the pattern for}%
\typeout{** the default language instead.}%
\else
\language=\csname l@#1\endcsname
\fi
#2}}
\providecommand{\BIBdecl}{\relax}
\BIBdecl

\bibitem{surveyEdgeShi}
H.~Xu, J.~Wu, Q.~Pan, X.~Guan, and M.~Guizani, ``A survey on digital twin for
  industrial internet of things: Applications, technologies and tools,''
  \emph{{IEEE} Commun. Surveys Tuts.}, vol.~25, no.~4, pp. 2569--2598, 2023.

\bibitem{denizMag}
D.~G{\"u}nd{\"u}z, D.~B. Kurka, M.~Jankowski, M.~M. Amiri, E.~Ozfatura, and
  S.~Sreekumar, ``Communicate to learn at the edge,'' \emph{{IEEE} Commun.
  Mag.}, vol.~58, no.~12, pp. 14--19, 2020.

\bibitem{ganjCascaded}
A.~Alabbasi, M.~Ganjalizadeh, K.~Vandikas, and M.~Petrova, ``On cascaded
  federated learning for multi-tier predictive models,'' in \emph{{IEEE} Int.
  Conf. Commun. Workshops (ICC Workshops)}, 2021.

\bibitem{dAIFaultDetectFactory}
H.~Huang \emph{et~al.}, ``Real-time fault detection for {IIoT} facilities using
  {GBRBM}-based {DNN},'' \emph{IEEE Internet Things J.}, vol.~7, no.~7, pp.
  5713--5722, 2020.

\bibitem{vehicularFlTsnm}
A.~Hammoud, H.~Otrok, A.~Mourad, and Z.~Dziong, ``On demand fog federations for
  horizontal federated learning in {IoV},'' \emph{{IEEE} Trans. Netw. Service
  Manag.}, vol.~19, no.~3, pp. 3062--3075, 2022.

\bibitem{fl}
B.~McMahan, E.~Moore, D.~Ramage, S.~Hampson, and B.~A.~y. Arcas,
  ``Communication-efficient learning of deep networks from decentralized
  data,'' in \emph{Proc. Int. Conf. Artif. Intell. Stat. (AISTATS)}, 2017.

\bibitem{duttaKsync}
S.~Dutta, G.~Joshi, S.~Ghosh, P.~Dube, and P.~Nagpurkar, ``Slow and stale
  gradients can win the race: Error-runtime trade-offs in distributed {SGD},''
  in \emph{Proc. Int. Conf. Artif. Intell. Stat. (AISTATS)}, 2018.

\bibitem{flHyperparamDevSelTsnm}
X.~Yu, Y.~Lin, Z.~Gao, H.~Du, and D.~Niyato, ``Dynamic and fast convergence for
  federated learning via optimized hyperparameters,'' \emph{{IEEE} Trans. Netw.
  Service Manag.}, early access, 2024.

\bibitem{chen2021Survey}
M.~Chen \emph{et~al.}, ``Distributed learning in wireless networks: Recent
  progress and future challenges,'' \emph{{IEEE} J. Sel. Areas Commun.},
  vol.~39, no.~12, pp. 3579--3605, 2021.

\bibitem{3GPP22104}
\textit{Service requirements for cyber-physical control applications in
  vertical domains}, 3GPP, TS 22.104 v19.2.0, 2024.

\bibitem{3GPP22261}
\textit{Service requirements for the 5G system}, 3GPP, TS 22.261 v20.1.0, 2025.

\bibitem{urllcApp}
O.~L.~A. López \emph{et~al.}, ``Statistical tools and methodologies for
  ultrareliable low-latency communication—a tutorial,'' \emph{Proc. {IEEE}},
  vol. 111, no.~11, pp. 1502--1543, 2023.

\bibitem{aiRanAlliance}
``{AI-RAN} alliance vision and mission,'' AI-RAN Alliance, White Paper, 2024.

\bibitem{urllcEmbbTransCoex}
A.~K. Bairagi \emph{et~al.}, ``Coexistence mechanism between {eMBB} and {uRLLC}
  in {5G} wireless networks,'' \emph{IEEE Trans. Commun.}, vol.~69, no.~3, pp.
  1736--1749, 2021.

\bibitem{embbUrllcTsnm}
U.~Bucci, D.~Cassioli, and A.~Marotta, ``Performance of spatially diverse
  {URLLC} and {eMBB} traffic in cell free massive {MIMO} environments,''
  \emph{{IEEE} Trans. Netw. Service Manag.}, vol.~21, no.~1, pp. 161--173,
  2024.

\bibitem{embbUrllcDistAiTsnm}
M.~Alsenwi, E.~Lagunas, and S.~Chatzinotas, ``Distributed learning framework
  for {eMBB}-{URLLC} multiplexing in open radio access networks,'' \emph{{IEEE}
  Trans. Netw. Service Manag.}, vol.~21, no.~5, pp. 5718--5732, 2024.

\bibitem{mimpUrllcMmtc}
L.~Valentini, E.~Bernardi, F.~Saggese, M.~Chiani, E.~Paolini, and P.~Popovski,
  ``Contention-based {mMTC}/{URLLC} coexistence through coded random access and
  massive {MIMO},'' \emph{{IEEE} J. Sel. Topics Signal Process.}, early access,
  2024.

\bibitem{popovski20185gUrllcEmbbMmtc}
P.~Popovski, K.~F. Trillingsgaard, O.~Simeone, and G.~Durisi, ``{5G} wireless
  network slicing for {eMBB}, {URLLC}, and {mMTC}: {A} communication-theoretic
  view,'' \emph{IEEE Access}, vol.~6, pp. 55\,765--55\,779, 2018.

\bibitem{ganjInterplay}
M.~Ganjalizadeh, H.~S. Ghadikolaei, J.~Haraldson, and M.~Petrova, ``Interplay
  between distributed {AI} workflow and {URLLC},'' in \emph{{IEEE} Global
  Commun. Conf. (GLOBECOM)}, 2022.

\bibitem{3GPP22874AiModel}
\textit{{5G} System ({5GS}); Study on traffic characteristics and performance
  requirements for {AI/ML} model transfer}, 3GPP, TR 22.874 v18.2.0, 2021.

\bibitem{3GPP22876AiModelP2}
\textit{Study on {AI}/{ML} Model Transfer Phase2}, 3GPP, TR 22.876 v19.1.0,
  2023.

\bibitem{Bottou2018SIAM}
L.~Bottou, F.~Curtis, and J.~Nocedal, ``Optimization methods for large-scale
  machine learning,'' \emph{SIAM Review}, vol.~60, no.~2, pp. 223--311, 2018.

\bibitem{scaman2018optimal}
K.~Scaman, F.~Bach, S.~Bubeck, L.~Massouli{\'e}, and Y.~T. Lee, ``Optimal
  algorithms for non-smooth distributed optimization in networks,'' \emph{Adv.
  Neural Inf. Process. Syst. (NeurIPS)}, vol.~31, 2018.

\bibitem{li2019federated}
T.~{Li}, A.~K. {Sahu}, A.~{Talwalkar}, and V.~{Smith}, ``Federated learning:
  {C}hallenges, methods, and future directions,'' \emph{IEEE Signal Process.
  Mag.}, vol.~37, no.~3, pp. 50--60, 2020.

\bibitem{ghadikolaei2021lena}
H.~S. Ghadikolaei, S.~Stich, and M.~Jaggi, ``{LENA}: Communication-efficient
  distributed learning with self-triggered gradient uploads,'' in \emph{Proc.
  Int. Conf. Artif. Intell. Stat. (AISTATS)}, 2021.

\bibitem{ji2020dynamic}
S.~Ji, W.~Jiang, A.~Walid, and X.~Li, ``Dynamic sampling and selective masking
  for communication-efficient federated learning,'' \emph{IEEE Intell. Syst.},
  vol.~37, no.~2, pp. 27--34, 2022.

\bibitem{chen2018lag}
T.~Chen, G.~Giannakis, T.~Sun, and W.~Yin, ``{LAG}: Lazily aggregated gradient
  for communication-efficient distributed learning,'' in \emph{Adv. Neural Inf.
  Process. Syst. (NeurIPS)}, 2018.

\bibitem{chenJointLearningCommFL}
M.~Chen, Z.~Yang, W.~Saad, C.~Yin, H.~V. Poor, and S.~Cui, ``A joint learning
  and communications framework for federated learning over wireless networks,''
  \emph{IEEE Trans. Wireless Commun.}, vol.~20, no.~1, pp. 269--283, 2021.

\bibitem{dinhFEDL}
C.~T. Dinh \emph{et~al.}, ``Federated learning over wireless networks:
  Convergence analysis and resource allocation,'' \emph{IEEE/ACM Trans. Netw.},
  vol.~29, no.~1, pp. 398--409, 2021.

\bibitem{energyDelayEarlyAccessTmcTd3}
W.~Hou \emph{et~al.}, ``Adaptive training and aggregation for federated
  learning in multi-tier computing networks,'' \emph{{IEEE} Trans. Mobile
  Comput.}, vol.~23, no.~5, pp. 4376--4388, 2024.

\bibitem{deviceSelMab}
W.~Xia, T.~Q.~S. Quek, K.~Guo, W.~Wen, H.~H. Yang, and H.~Zhu, ``Multi-armed
  bandit-based client scheduling for federated learning,'' \emph{{IEEE} Trans.
  Wireless Commun.}, vol.~19, no.~11, pp. 7108--7123, 2020.

\bibitem{energyDelayTiiDdpg}
P.~Zhang, C.~Wang, C.~Jiang, and Z.~Han, ``Deep reinforcement learning assisted
  federated learning algorithm for data management of {IIoT},'' \emph{{IEEE}
  Trans. Ind. Informat.}, vol.~17, no.~12, pp. 8475--8484, 2021.

\bibitem{optFlIIoT}
W.~Zhang \emph{et~al.}, ``Optimizing federated learning in distributed
  industrial {IoT}: A multi-agent approach,'' \emph{IEEE J. Sel. Areas
  Commun.}, vol.~39, no.~12, pp. 3688--3703, 2021.

\bibitem{denizUpdateAware}
M.~M. Amiri, D.~G{\"u}nd{\"u}z, S.~R. Kulkarni, and H.~V. Poor, ``Convergence
  of update aware device scheduling for federated learning at the wireless
  edge,'' \emph{IEEE Trans. Wireless Commun.}, vol.~20, no.~6, pp. 3643--3658,
  2021.

\bibitem{denizSpawc}
M.~E. Ozfatura, J.~Zhao, and D.~G{\"u}nd{\"u}z, ``Fast federated edge learning
  with overlapped communication and computation and channel-aware fair client
  scheduling,'' in \emph{{IEEE} Int. Workshop Signal Process. Adv. Wireless
  Commun. (SPAWC)}, 2021.

\bibitem{denizAmiri}
M.~M. Amiri, D.~G{\"u}nd{\"u}z, S.~R. Kulkarni, and H.~V. Poor, ``Convergence
  of federated learning over a noisy downlink,'' \emph{{IEEE} Trans. Wireless
  Commun.}, vol.~21, no.~3, pp. 1422--1437, 2022.

\bibitem{flFairSelTsnm}
A.~M. Albaseer, M.~Abdallah, A.~Al-Fuqaha, A.~M. Seid, A.~Erbad, and O.~A.
  Dobre, ``Fair selection of edge nodes to participate in clustered federated
  multitask learning,'' \emph{{IEEE} Trans. Netw. Service Manag.}, vol.~20,
  no.~2, pp. 1502--1516, 2023.

\bibitem{dahlman5GNr}
E.~Dahlman, S.~Parkvall, and J.~Skold, \emph{{5G NR}: The next generation
  wireless access technology}, 2nd~ed.\hskip 1em plus 0.5em minus 0.4em\relax
  New York, NY, USA: Academic, 2020.

\bibitem{3GPP38901}
\textit{Study on channel model for frequencies from 0.5 to 100 {GHz}}, 3GPP, TR
  38.901 v18.0.0, 2024.

\bibitem{5gAcia}
``Service-level specifications {(SLSs)} for {5G} technology-enabled connected
  industries,'' 5G Alliance for Connected Industries and Automation
  ({5G-ACIA}), Frankfurt, Germany, White Paper, 2021.

\bibitem{ganjPimrcTranslation}
M.~Ganjalizadeh, A.~Alabbasi, J.~Sachs, and M.~Petrova, ``Translating
  cyber-physical control application requirements to network level
  parameters,'' in \emph{{IEEE} Int. Symp. Pers., Indoor, Mobile Radio Commun.
  (PIMRC)}, 2020.

\bibitem{ganjTIIOrch}
M.~Ganjalizadeh, H.~S. Ghadikolaei, A.~Azari, A.~Alabbasi, and M.~Petrova,
  ``Saving energy and spectrum in enabling {URLLC} services: A scalable {RL}
  solution,'' \emph{{IEEE} Trans. Ind. Informat.}, vol.~19, no.~10, pp.
  10\,265--10\,276, Oct. 2023.

\bibitem{popovskiUrllc}
P.~Popovski \emph{et~al.}, ``Wireless access in ultra-reliable low-latency
  communication {(URLLC)},'' \emph{IEEE Trans. Commun.}, vol.~67, no.~8, pp.
  5783--5801, 2019.

\bibitem{nonconvexOptBehrouz}
T.~Tatarenko and B.~Touri, ``Non-convex distributed optimization,''
  \emph{{IEEE} Trans. Autom. Control}, vol.~62, no.~8, pp. 3744--3757, 2017.

\bibitem{sac}
T.~Haarnoja \emph{et~al.}, ``Soft actor-critic algorithms and applications,''
  \emph{\textup{2018}, arXiv:1812.05905 \textup{[cs.LG]}}.

\bibitem{URLLCTRS}
M.~Bennis, M.~Debbah, and H.~V. Poor, ``Ultrareliable and low-latency wireless
  communication: Tail, risk, and scale,'' \emph{Proc. {IEEE}}, vol. 106,
  no.~10, pp. 1834--1853, Oct. 2018.

\bibitem{tesslerreward}
C.~Tessler, D.~J. Mankowitz, and S.~Mannor, ``Reward constrained policy
  optimization,'' in \emph{Int. Conf. Learn. Represent. (ICLR)}, 2019.

\bibitem{donti2021dcreward}
P.~L. Donti, D.~Rolnick, and J.~Z. Kolter, ``{DC}3: A learning method for
  optimization with hard constraints,'' in \emph{Int. Conf. Learn. Represent.
  (ICLR)}, 2021.

\bibitem{td3}
S.~Fujimoto, H.~van Hoof, and D.~Meger, ``Addressing function approximation
  error in actor-critic methods,'' in \emph{Proc. Int. Conf. Mach. Learn.
  (ICML)}, 2018.

\bibitem{mobileNets}
A.~G. Howard \emph{et~al.}, ``{MobileNets}: Efficient convolutional neural
  networks for mobile vision applications,'' \emph{\textup{2017},
  arXiv:1704.04861 \textup{[cs.CV]}}.

\bibitem{Sindri2020Linear}
S.~Magnusson, H.~S.~Ghadikolaei, and N.~Li, ``On maintaining linear convergence
  of distributed learning and optimization under limited communication,''
  \emph{IEEE Trans. Signal Process.}, vol.~68, pp. 6101--6116, 2020.

\bibitem{prioReplay}
T.~Schaul, J.~Quan, I.~Antonoglou, and D.~Silver, ``{Prioritized experience
  replay},'' in \emph{Int. Conf. Learn. Represent. (ICLR)}, 2016.

\bibitem{confBound}
E.~L. Kaplan and P.~Meier, ``Nonparametric estimation from incomplete
  observations,'' \emph{J. Amer. Statistical Assoc.}, vol.~53, no. 282, pp.
  457--481, 1958.

\bibitem{gradientNoise}
M.~Cohen, J.~Diakonikolas, and L.~Orecchia, ``On acceleration with
  noise-corrupted gradients,'' in \emph{Proc. Int. Conf. Mach. Learn. (ICML)},
  2018.

\bibitem{flConvergNoniid}
X.~Li, K.~Huang, W.~Yang, S.~Wang, and Z.~Zhang, ``On the convergence of
  {FedAvg} on non-{IID} data,'' in \emph{Int. Conf. Learn. Represent. (ICLR)},
  2020.

\end{thebibliography}
\clearpage

\end{document}